\documentclass[12pt]{article}
\usepackage[top=1in, bottom=1in, left=1in, right=1in]{geometry}                
\usepackage{graphicx}
\usepackage{bm,amssymb,amsmath,amsthm}
\usepackage{epstopdf}
\usepackage{verbatim}
\usepackage{multirow}
\usepackage{setspace}
\usepackage[flushleft]{threeparttable}
\usepackage{textcomp}
\usepackage{tikz,makecell}

\usetikzlibrary{positioning}
\usepackage{hyperref}
\usepackage{color}
\usepackage{latexsym,graphicx,color}
\usepackage[natbibapa]{apacite}
\usepackage{url}

\newtheorem{procedure}{Procedure}

\title{Latent Feature Extraction for Process Data via Multidimensional Scaling}
\author{Xueying Tang, Zhi Wang, Qiwei He, Jingchen Liu, and Zhiliang Ying}
\date{}

\begin{document}
\maketitle
\baselineskip 18pt
\begin{abstract}
Computer-based interactive items have become prevalent in recent educational assessments. In such items, the entire human-computer interactive process is recorded in a log file and is known as the response process.
This paper aims at extracting useful information from response processes. 
In particular, we consider an exploratory latent variable analysis for process data. 
Latent variables are extracted through a multidimensional scaling framework and can be empirically proved to contain more information than classic binary responses in terms of out-of-sample prediction of many variables.
\end{abstract}

\section{Introduction}\label{sec:intro}

Computer-based problem-solving items have become prevalent in large-scale assessments. These items are developed to measure skills related to problem solving in work and personal life. Thanks to the human-computer interface, it is possible to record the entire problem-solving process, as is the case of scientific inquiry items in the Programme for International Student Assessment (PISA) and Problem Solving in Technology-Rich Environments (PSTRE) items in the Programme for the International Assessment of Adult Competencies (PIAAC). 
The responses of such items are complex and are often in the form of a process.
More precisely, the record of each item response contains a sequence of ordered and time-stamped actions.

\begin{figure}[htb]
\includegraphics[width=\textwidth]{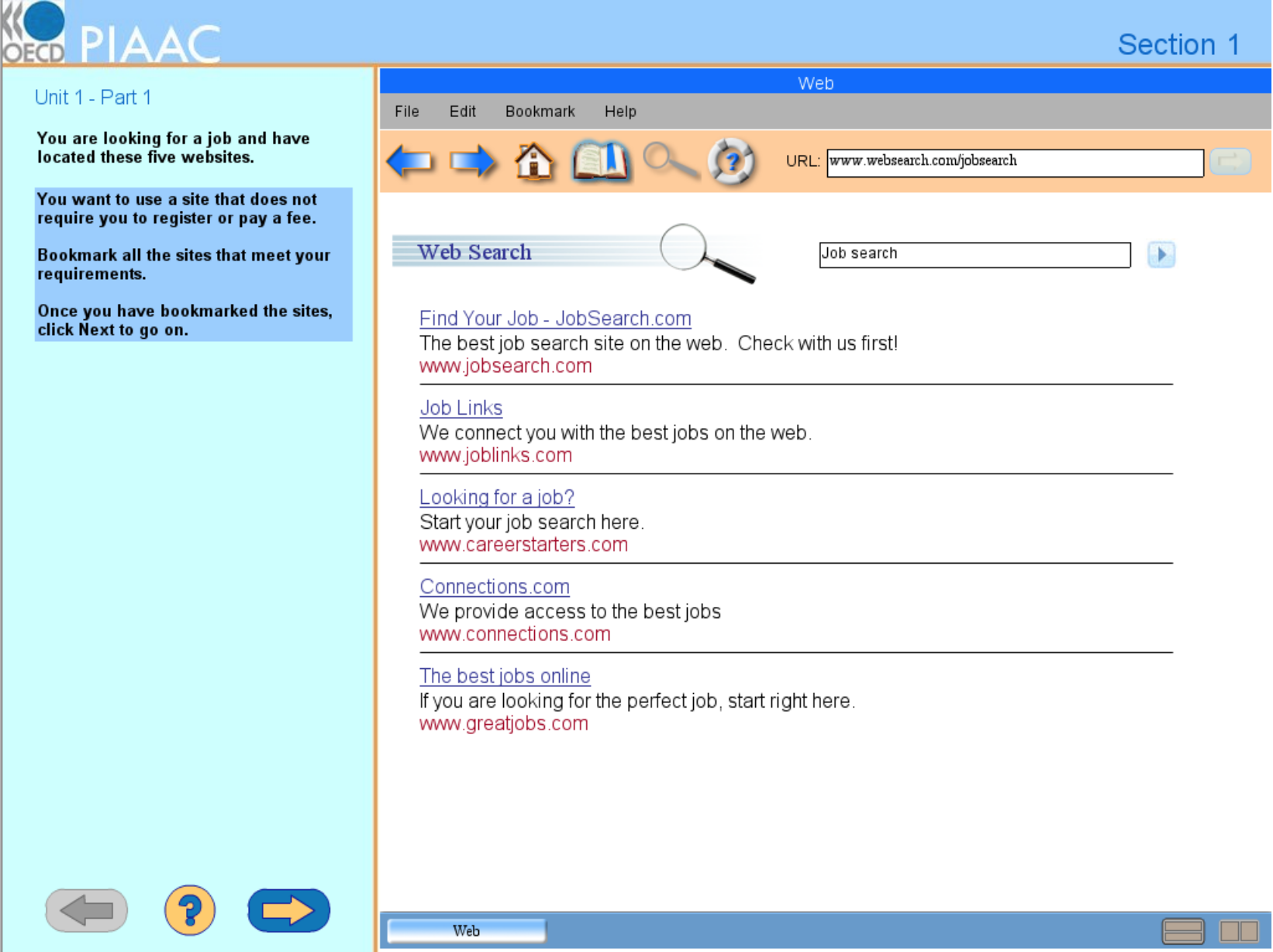}
\caption{Main page of the sample item.}\label{fig:item_main}
\end{figure}

\begin{figure}[htb]
\includegraphics[width=\textwidth]{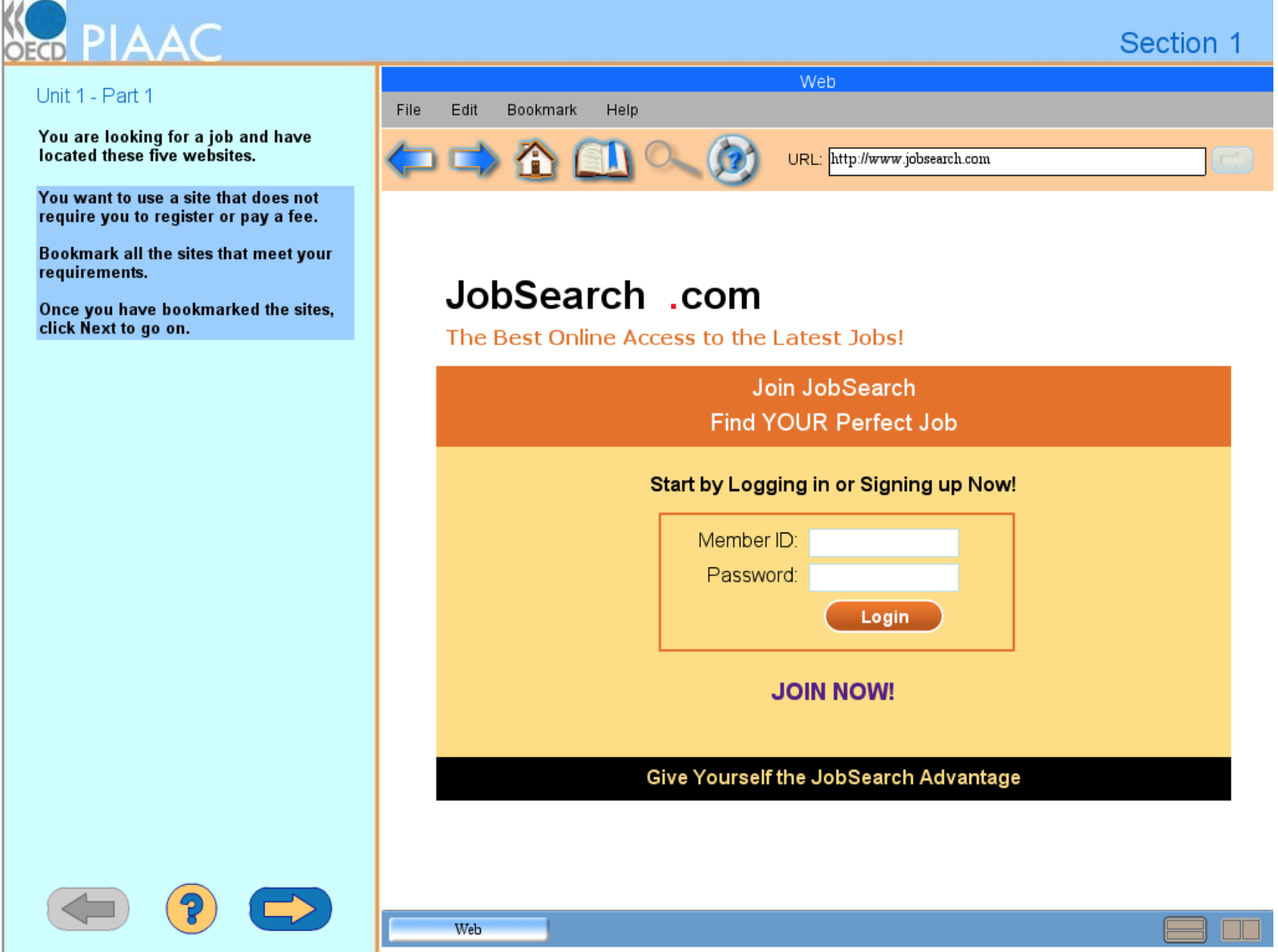}
\caption{Website in the first link in Figure \ref{fig:item_main}.}\label{fig:item_page1}
\end{figure}

\begin{figure}[htb]
\includegraphics[width=\textwidth]{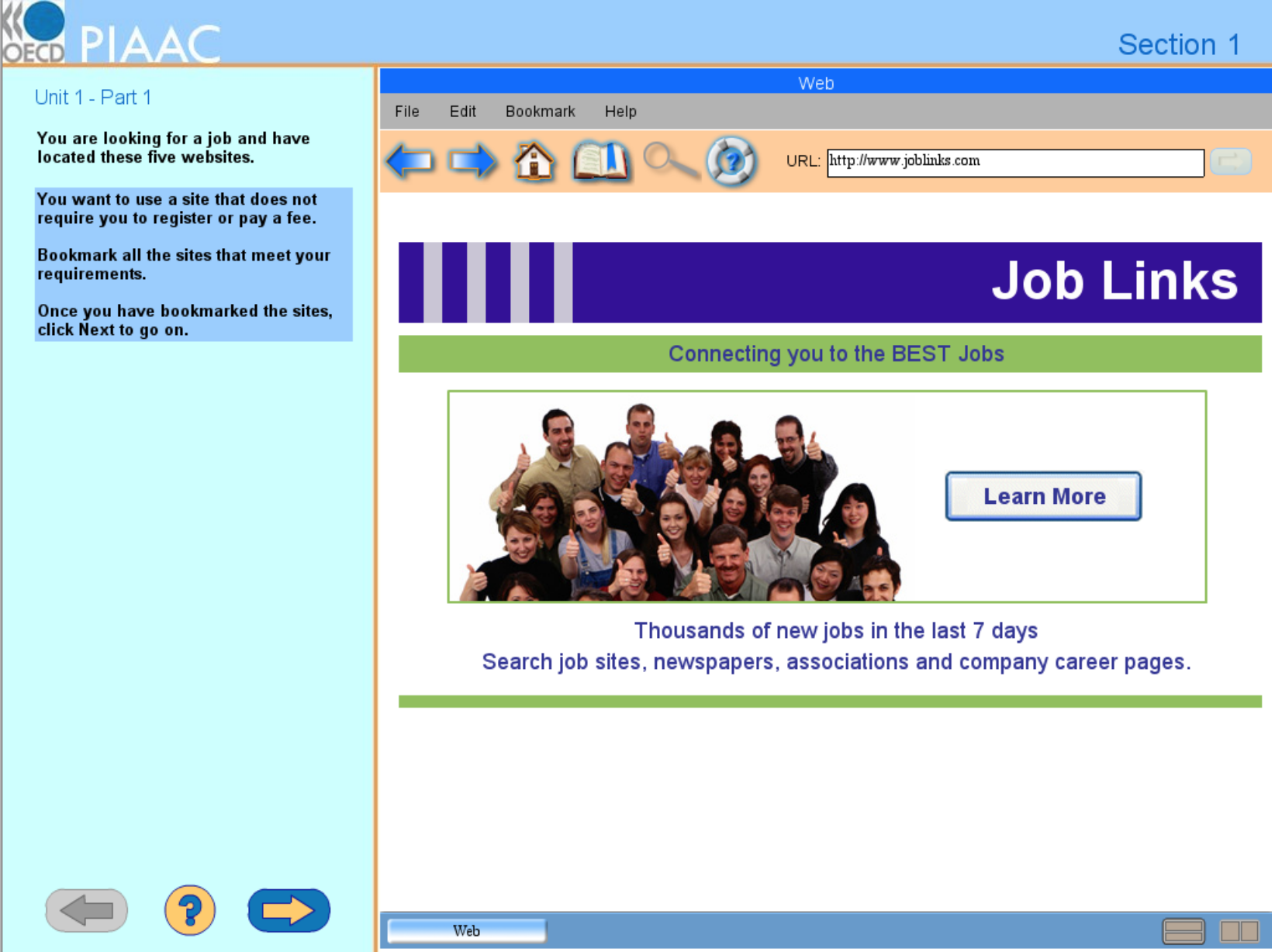}
\caption{Website in the second link in Figure \ref{fig:item_main}.}\label{fig:item_page2}
\end{figure}

An example of a PIAAC PSTRE item is shown in Figures \ref{fig:item_main}--\ref{fig:item_page2}. Figure \ref{fig:item_main} displays the main page of the item. The left panel of the main page provides item instructions. In this item, respondents are asked to identify websites that do not require registration or a fee from those listed in the web browser in the right panel. Respondents can visit a website by clicking its link. Figures \ref{fig:item_page1} and \ref{fig:item_page2} show the web pages of the first and the second links, respectively. Further information of the second website can be found by clicking on the ``Learn More'' button shown in Figure \ref{fig:item_page2}. If a website is considered useful, it can be bookmarked by either using the menu item ``Bookmark'' or clicking the bookmark icon in the tool bar. Suppose that a respondent completes the task by clicking on the first link, reading the first website, going back to the main page, clicking on the second link, and bookmarking the second website by clicking the bookmark icon. All these actions are recorded in the log file in order. The sequence ``Start, Click\_W1, Back, Click\_W2, Toolbar\_Bookmark, Next'' constitutes a response process. 

In this paper, we present a generic method to extract useful information regarding participants from their response processes.
Latent variable or latent class models have been used in the literature to summarize item responses. 
Existing models and methods are not directly applicable to response processes. 
The analysis of process data is difficult for several reasons.
First, response processes are in a nonstandard format. A response process is a sequence of actions and each action is a categorical variable. In addition, process length varies across individuals. Because of the nonstandard format, classic latent variable models such as item response theory models \citep{lord1980applications} and cognitive diagnosis models \citep{rupp2010diagnostic} do not apply to process data.
Second, computer-based assessments and their log files cover a large variety of items. Every human-computer interface generates log files. This makes confirmatory analysis practically infeasible due to the large amount and variety of items. It is too expensive to perform confirmatory analysis for each potential human-computer interface and then verify it empirically. Furthermore, the cognitive process of human-computer interaction is not thoroughly understood, which adds to the difficulty of confirmatory analysis.
Lastly, response processes are often very noisy. For instance, the lagged correlations of action occurrences are often very close to zero, that is, response processes behave like white noise from an autoregressive process viewpoint. 

Assessment of data beyond traditional responses have been studied previously. It has been shown that item response time can reveal test-taker response behaviors that are helpful for test design \citep{van2008using,qian2016using}. Models have been proposed to perform cognitive assessments using both traditional responses and response time \citep{entink2009multivariate,wang2018using,zhan2018cognitive}. The study of process data is at a more preliminary stage. Most works such as \citet{greiff2016understanding} and \citet{kroehne2018conceptualize} first summarized process data into several variables and then investigated their relationship with other variables of interest using standard statistical methods. The design of these summary variables is usually item specific and thus hard to generalize. \citet{he2015identifying,he2016analyzing} explored the association between action sequence patterns and traditional responses using n-grams. Although the procedure of extracting n-gram features is generic, the sequence patterns under consideration are limited.

The objective of the present analysis is to perform exploratory analysis on process data. In particular, we propose a generic method to extract features (latent variables) from response processes. The proposed method does not rely on prior knowledge of the items or the response processes and is applicable essentially to all process responses. We apply it to all 14 PIAAC PSTRE items that cover a range of human-computer interfaces.

The basic technique of our proposed method is multidimensional scaling \citep{borg2005modern}. It constructs features based on the relative differences among individuals. 
Though numerous variants of multidimensional scaling (MDS) exist, their common goal is to locate objects in a vector space according to their pairwise dissimilarities in such a way that similar objects are close together, while less similar objects are far apart. MDS has been used for data visualization and dimension reduction in cognitive diagnosis, test analysis, and many other areas of psychometrics \citep{skager1966multidimensional,karni1972use,subkoviak1975use,shoben1983applications,meyer2018scores}. In the context of process data analysis, if the differences between two processes can be properly summarized by a dissimilarity measure, then the coordinates obtained from MDS can be treated as features storing information of the original processes. With a proper rotation, each feature describes the variation of certain ability or behavior pattern among the group of respondents.


We use a prediction procedure to demonstrate that response processes contain more information than traditional item responses. We denote the features extracted from response processes by $\theta$. For each response process, there is a binary response, denoted by $r$, indicating whether the respondent has successfully accomplished the task. 
To compare the information contained in $\theta$ and $r$, we adopt a third variable, denoted by $y$ (such as numeracy score, literacy score, etc.), and inspect the prediction of $y$ based on $r$ and that based on $\theta$. In the empirical analysis of PSTRE in PIAAC, we find that the prediction based on $\theta$ outperforms that based on $r$ for a wide range of $y$ variables including assessment scores, basic demographic variables, and some background variables. 

The rest of this paper is organized as follows. In Section \ref{sec:mds_feature}, we introduce a dissimilarity measure for action sequences and describe the proposed feature extraction procedure. A simulation study is presented in Section \ref{sec:simulation} to demonstrate the procedure and how the latent structure of action sequences are reflected in extracted features. In Section \ref{sec:example}, we show through a case study of PIAAC PSTRE item response processes that features extracted from process data contain much richer information than binary responses. Section \ref{sec:discussion} contains some concluding remarks.




\section{Feature Extraction via Multidimensional Scaling}\label{sec:mds_feature}


Consider a problem-solving item in which a student takes a number of actions to complete a task. We use  $\mathcal{A}=\{a_1, \ldots, a_N\}$ to denote the set of possible actions of this item where $N$ is the number of distinct actions.
A response process is a sequence of actions $\bm s = (s_1, \ldots, s_{L})$ where each $s_i$ is an action in $\mathcal{A}$ and $L$ is the process length, i.e., the number of actions taken in the response process.
An action in $\mathcal{A}$ may appear multiple times or never appear in $\bm s$. 
We observed the response processes of $n$ students and use subscript to index different observations: $\bm s_1, \ldots, \bm s_n$.
The process length also varies among individuals; we use $L_i$ to denote the length of ${\bm s}_i$.
The heterogeneous length of response processes for the same item is one of the technical difficulties in process data analysis.
In what follows, we describe a procedure that transforms the response processes with heterogeneous length to homogeneous-dimension latent vectors that may be used for standard analysis.

The core of the procedure is MDS, which has been widely used as a data visualization and dimension reduction tool in many fields including psychometrics \citep{takane200611}. The goal of MDS is to locate objects in a vector space according to their pairwise dissimilarities in such a way that similar objects are close together, while dissimilar objects are far apart. 
We begin the discussion with a description of a dissimilarity measure between discrete action sequences.
This measure is key to the subsequent application of multidimensional scaling and it summarizes the variation among response processes.
An appropriate dissimilarity measure should accommodate three characteristics of response processes. 
First, process data is a collection of discrete processes on which arithmetic calculation can not be performed. 
Second, processes from different respondents are of very different lengths. 
Third, the order of actions matters. 
Although the order of actions may not affect the final outcome of the task, it reflects respondents' habit and personality.

Based on these considerations, we adopt the following dissimilarity measure. Let $\bm s_i = (s_{i1}, \ldots, s_{iL_i})$ and $\bm s_j = (s_{j1}, \ldots, s_{jL_j})$ be two action sequences.
Define the dissimilarity between $\bm s_i$ and $\bm s_j$ as
\begin{equation}\label{eq:dissimilarity}
d(\bm s_i, \bm s_j) = \frac{f(\bm s_i, \bm s_j) + g(\bm s_i, \bm s_j)}{L_i + L_j},
\end{equation}
where $f(\bm s_i, \bm s_j)$ quantifies the dissimilarity among the actions that appear in both $\bm s_i$ and $\bm s_j$ and $g(\bm s_i, \bm s_j)$ is the count of  actions appearing in only one of $\bm s_i$ and $\bm s_j$.

We now provide the precise definition of $f$ and $g$. 
For an action $a \in \mathcal{A}$, let $\bm s^a$ be a sequence consisting of chronologically ordered positions of $a$ in sequence $\bm s$. 
The length of $\bm s^a$, $L^a$, is the number of times that $a$ appears in $\bm s$.
We use $\bm s^a(k)$ to denote the $k$th element of $\bm s^a$, namely, the position of the $k$th appearance of  $a$ in $\bm s$. 
For two sequences $\bm s_i$ and $\bm s_j$, let $C_{ij}$ denote the set of actions that appear in both $\bm s_i$ and $\bm s_j$ and $U_{ij}$ denote the set of actions that appear in $\bm s_i$ but not in $\bm s_j$. Then $f(\bm s_i, \bm s_j)$ and $g(\bm s_i, \bm s_j)$ are defined as
\begin{equation}\label{eq:dissimilarity_f}
f(\bm s_i, \bm s_j) = \frac{\sum_{a \in C_{ij}} \sum_{k=1}^{K_{ij}^a} | \bm s_i^a(k) - \bm s_j^a(k)|}{\max\{ L_i, L_j \}},
\end{equation}
and 
\begin{equation}\label{eq:dissimilarity_g}
g(\bm s_i, \bm s_j) = \sum_{a \in U_{ij}} L_i^a + \sum_{a \in U_{ji}} L_j^a,
\end{equation}
where $K_{ij}^a= \min (L_i^a, L_j^a)$. This dissimilarity measure is first proposed in \citet{gomez2008similarity} to measure differences in tourists' itineraries and differences in websites visited by certain users.

We use a simple example to demonstrate how the dissimilarity is calculated. Consider a set of four possible actions $\mathcal{A} = \{X, Y, Z, W\}$ and two sequences, $\bm s_1 = (X, Y, X, Y, Z)$ and $\bm s_2 = (W, X, Y, W)$. Since $X$ and $Y$ appear in both sequences, $C_{ij} = \{X, Y\}$. Action $X$ appears in $\bm s_1$ at positions 1 and 3 and appears in $\bm s_2$ in position 2, so $\bm s^X_1 = (1, 3)$ and $\bm s^X_2 = (2)$. The difference between $\bm s_1$ and $\bm s_2$ in the appearance of $X$ is $|1 - 2| = 1$. Similarly, we can find $\bm s^Y_1 = (2, 4)$, $\bm s^Y_2 = (3)$ and the difference in the appearance of $Y$ is $|2 - 3| = 1$. Therefore, $f(\bm s_1, \bm s_2) = (|1 - 2| + | 2 - 3|) / 5= 0.4$. Since $U_{21} = \{W\}$ and $U_{12} = \{Z\}$ with $W$ appearing twice in $\bm s_2$ and $Z$ appearing once in $\bm s_1$, $g(\bm s_1, \bm s_2) = 2 + 1 = 3$. According to \eqref{eq:dissimilarity}, $d(\bm s_1, \bm s_2) = (0.4 + 3 )/9 = 0.38$.

The calculation of the dissimilarity described in \eqref{eq:dissimilarity} does not require inputs of informative behavior patterns or the meaning of each action. This is crucial for our automated feature extraction procedure at the exploratory stage of the analysis. 

For action sequences $\bm s_1, \ldots, \bm s_n$, let an $n \times n$ symmetric matrix $\bm D = (d_{ij})$ denote their dissimilarity matrix, where $d_{ij} = d(\bm s_i, \bm s_j)$ measures the dissimilarity between $\bm s_i$ and $\bm s_j$, $i,j= 1, \ldots, n$. 
Higher dissimilarities indicate larger differences and the dissimilarity between two identical objects is zero, namely, $d_{ii} = 0$ for $i=1, \ldots, n$. 
MDS maps each action sequence to a latent vector $\bm x$ in the $K$-dimensional Euclidean space $\mathbb{R}^K$ such that they govern the dissimilarities.
Mathematically, applying MDS to objects with dissimilarity matrix $\bm D$ essentially minimizes 
\begin{equation}\label{eq:mds_obj}
\sum_{i<j} \left(d_{ij} - \|\bm x_i - \bm x_j\|\right)^2
\end{equation}
with respect to $\bm X = (\bm x_1, \ldots, \bm x_n)^T$, where $\bm x_i \in \mathbb{R}^K$ is the latent vector of $\bm s_i$ in $\mathbb{R}^K$ and $\| \bm x_i - \bm x_j\| = \sqrt{(\bm x_i - \bm x_j)^T (\bm x_i - \bm x_j)}$. Many algorithms have been proposed to solve the optimization problem. For simplicity, we use stochastic gradient descent \citep{robbins1951stochastic} to minimize \eqref{eq:mds_obj}.


Combining the calculation of the dissimilarity matrix and MDS, we present the feature extraction procedure for process data.

\begin{procedure}[Feature extraction for process data]\label{proc:mds_feature} 
~
\begin{enumerate}
\item Compute the dissimilarity matrix $\bm D$ of $n$ action sequences $\bm s_1, \bm s_2, \ldots, \bm s_n$ by calculating the pairwise dissimilarities $d_{ij}, 1\leq i, j \leq n$ according to \eqref{eq:dissimilarity}, \eqref{eq:dissimilarity_f} and \eqref{eq:dissimilarity_g}.
\item Obtain $K$ raw features $\tilde{\bm x}_1, \ldots, \tilde{\bm x}_K$ by minimizing \eqref{eq:mds_obj}.
\item Obtain $K$ principal features $\bm x_1, \ldots, \bm x_K$ by performing principal component analysis (PCA) on the $K$ raw features.
\end{enumerate}
\end{procedure}

Procedure \ref{proc:mds_feature} extracts features with homogeneous dimension from action sequences with heterogeneous length. These features have a standard form and, as we will show in the simulation and case study, contain compressed information of the original sequences. Therefore, they can be easily incorporated as a surrogate of the action sequences in well-developed statistical models such as (generalized) linear models to study how process data reflects respondents' latent traits and how it is related to other quantities of interest. We will demonstrate how these can be achieved in the next two sections.

Principal component analysis is performed in Step 3 of Procedure \ref{proc:mds_feature} mainly for seeking interpretations of the features. As we will show in the case study, the first several principal features usually have clear interpretations, although the feature extraction procedure does not take into account the meaning of actions.

Procedure \ref{proc:mds_feature} requires the specification of $K$, the number of features to be extracted. If $K$ is too small, there are not enough features to characterize the variation of action sequences, leading to substantial information loss in extracted features. On the other hand, if $K$ is too large, some features can be redundant and can cause overfitting and instability in subsequent analyses. A suitable $K$ can be chosen by $m$-fold cross-validation. We randomly split the $n(n-1)/2$ pairwise dissimilarities into $m$ subsets. For each candidate value of $K$ and each subset of dissimilarities, we perform MDS using the rest of dissimilarities and calculating the discrepancy between the estimated and true dissimilarities for the subset. The value of $K$ that produces the smallest total discrepancy among $m$ subsets is chosen as the number of features to be extracted. This cross-validation procedure is summarized as follows.
\begin{procedure}[Choose $K$ by cross-validation]\label{proc:chooseK} 
~
\begin{enumerate}
\item Randomly split $\Omega=\{(i, j) : i < j; i, j = 1, \ldots, n\}$ into $m$ subsets $\Omega_1, \Omega_2, \ldots, \Omega_m$.
\item For each candidate value of $K$ and each $q$ in $\{1, 2 , \ldots, m\}$, obtain $\bm x_i^{(K, q)}, i = 1, \ldots, n$, by minimizing $$\sum_{(i, j) \in \Omega_{(-q)}} \left(d_{ij} - \|\bm x_i - \bm x_j\|\right)^2$$ with respect to $\bm x_1, \ldots, \bm x_n,$ where $\Omega_{(-q)} = \Omega \setminus \Omega_{q}$.
\item For each candidate value of $K$, calculate $$V(K) = \sum_{q=1}^m\sum_{(i, j) \in \Omega_{q}} \left(d_{ij} - \|\bm x_i^{(K, q)} - \bm x_j^{(K, q)}\|\right)^2.$$
\item Choose $K$ that produces the smallest $V(K)$.
\end{enumerate}
\end{procedure}

\section{Simulations}\label{sec:simulation}
In this section, we demonstrate the proposed feature extraction procedure on simulated data. 

\subsection{Data Generation}
Twenty-six possible actions ($N=26$) are considered in our simulations. Each possible action is denoted by an upper-case English letter, namely, $\mathcal{A} = \{\text{A, B, \ldots, Z}\}$ with $a_1 = \text{A}$ and $a_{N} = \text{Z}$. We use A and Z to denote the start and the end of an item. As a result, each action sequence starts with A and ends with Z.

The action sequences used in this section are generated from a Markov model, that is characterized by a probability transition matrix $\mathbf{P} = (p_{ij})_{1\leq i,j \leq N}$, whose element in the $i$th row and $j$th column is the probability that the next action is $a_j$ given the current action is $a_i$, where 
$P(s_{t+1} = a_j \, | \, s_t = a_i) = p_{ij}$. Because of the special roles of A and Z, the first element in each row of $\mathbf{P}$ is zero and all the elements in the last row except for the last one are zeros. Therefore, the Markov model for generating action sequences is determined by the $(N-1) \times (N-1)$ submatrix in the upper right corner of $\mathbf{P}$. We call this submatrix the core matrix of $\mathbf{P}$ and denote it by $\tilde{\mathbf{P}}$. The probability transition matrices used in our simulation study are randomly generated. The way in which they are generated will be explained in detail in the experiment settings. Given a probability transition matrix $\mathbf{P}$, we generate an action sequence by starting from A and sampling the subsequent actions according to $\mathbf{P}$ until Z appears.

\subsection{Experiment Settings}
We consider two strategies for generating action sequences. 
With strategy I, a set of $n$ action sequences are generated from the previous Markov model under two different transition matrices, $n/2$ sequences for each matrix. Action sequences generated from this strategy have a latent group structure. Sequences generated from the same transition matrix form a group and tend to be similar. 
The two probability transition matrices used in this strategy are randomly generated. 
Both of the matrices are generated by first constructing an $(N-1) \times (N-1)$ matrix $\mathbf{U}$. The elements of $\mathbf{U}$ are generated independently from a uniform distribution on interval $[-10, 10]$. Then $\tilde{\mathbf{P}} = (\tilde{p}_{ij})_{1 \leq i, j \leq N-1}$ is computed from $\mathbf{U}$ by
\begin{equation}
\tilde{p}_{ij} = \frac{\exp(u_{ij})} { \sum_{l=1}^{N-1}\exp(u_{il})}.
\end{equation}
In strategy II, each of $n$ action sequences is generated from a unique probability transition matrix. To construct these matrices, we first obtain a uniform matrix $\mathbf{U}$ as in strategy I. Then we draw $n$ independent samples, $\theta_0^{(1)}, \ldots,\theta_0^{(n)}$, from $N(0, 4)$ and compute the core matrix $\tilde{\mathbf P}^{(i)}$ for the $i$th sequence according to 
\begin{equation}
\tilde{p}_{jk}^{(i)} = \frac{\exp(\theta_0^{(i)}u_{jk})} { \sum_{l=1}^{N-1}\exp(\theta_0^{(i)}u_{jl})}.
\end{equation}
With this strategy, sequences with similar $\theta_0$ resemble each other. In other words, $\theta_0$ serves as a continuous latent variable determining the characteristics of the sequences. 

 We consider three choices of $n$, 200, 500, and 1000. For each strategy and each choice of $n$, we generate 100 sets of action sequences and extract features according to Procedure \ref{proc:mds_feature}. The number of features to be extracted are chosen by five-fold cross-validation described in Procedure \ref{proc:chooseK}.
 
To show that extracted features retain the information in action sequences, we derive several variables from action sequences for each dataset and examine how well these derived variables can be reconstructed from the extracted features. Good reconstruction performances indicate that a significant amount of information in action sequences is preserved in extracted features. The derived variables are indicators describing whether a unigram or a bigram appears in a sequence. We do not consider indicators for unigrams and bigrams that appears fewer than $0.05n$ times in a dataset. Logistic regression is used to reconstruct the derived variables from extracted features. For each data set, $n$ sequences are split into training and test sets in the ratio of 4:1. A logistic regression model is estimated for each derived variable on the training set and its prediction performance is evaluated on the test set. The average prediction accuracy and the worst prediction accuracy among all the derived variables are recorded for each dataset.

To inspect the ability of the extracted features in unveiling the latent structures in action sequences, we build a logistic regression model to identify the group structure from the extracted features for datasets generated from strategy I and a linear regression model of $\theta_0$ on the extracted features for datasets generated from strategy II. The models are fitted on the training set. The logistic model of group identity is evaluated by the prediction accuracy on the test set while the linear regression model of $\theta_0$ is evaluated by out-of-sample $R^2$ ($\text{OSR}^2$), the square of the correlation between the predicted and true values. As an analogy to the in-sample $R^2$ in linear regression, a higher $\text{OSR}^2$ indicates a better prediction performance.

\subsection{Results}

\begin{figure}[htb]
\centering
\includegraphics[width=\textwidth]{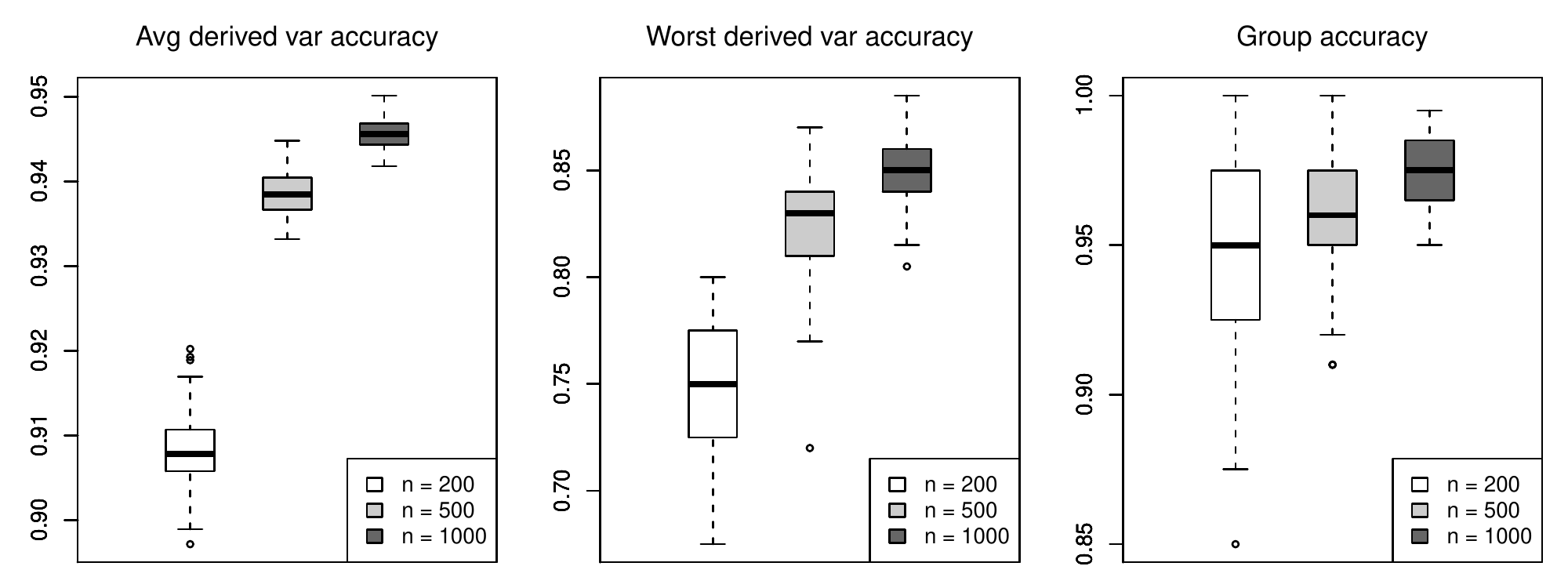}
\caption{Simulation results for datasets generated from strategy I.}\label{fig:box_sim_ex1}
\end{figure}

\begin{figure}[htb]
\centering
\includegraphics[width=\textwidth]{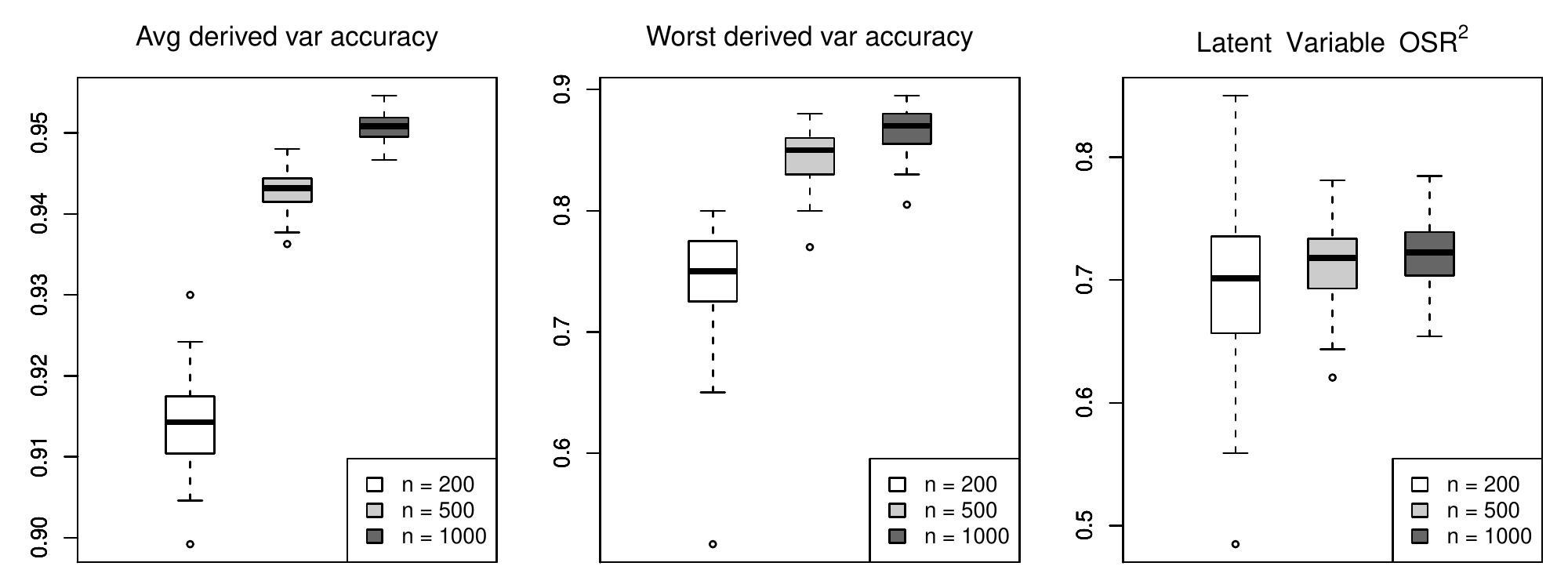}
\caption{Simulation results for datasets generated from strategy II.}\label{fig:box_sim_ex2}
\end{figure}

Figures \ref{fig:box_sim_ex1} and \ref{fig:box_sim_ex2} display the results for datasets generated by strategies I and II, respectively. The left and middle panels of both figures present the average and worst prediction accuracy for derived variables. Under all the settings, for almost all datasets, the averaged prediction accuracy is greater than 0.9 and the worst prediction accuracy is greater than 0.7. These results demonstrate that the derived variables can be reconstructed well and imply that a significant amount of information in action sequences is compressed into the extracted features. 

The right panel of Figure \ref{fig:box_sim_ex1} presents the prediction accuracy for group identity. For most of the datasets, the prediction accuracy is higher than 0.9, indicating that group structures in action sequences can be identified very accurately by extracted features. The right panel of Figure \ref{fig:box_sim_ex2} gives the $\text{OSR}^2$ for predicting $\theta_0$. It reflects that continuous latent characteristics in action sequences can be captured well by features extracted from Procedure \ref{proc:mds_feature} as the correlation between the predicted and true values is higher than 0.8 for most of the datasets. 


\begin{figure}[htb]
\centering
\includegraphics[width=0.45\textwidth]{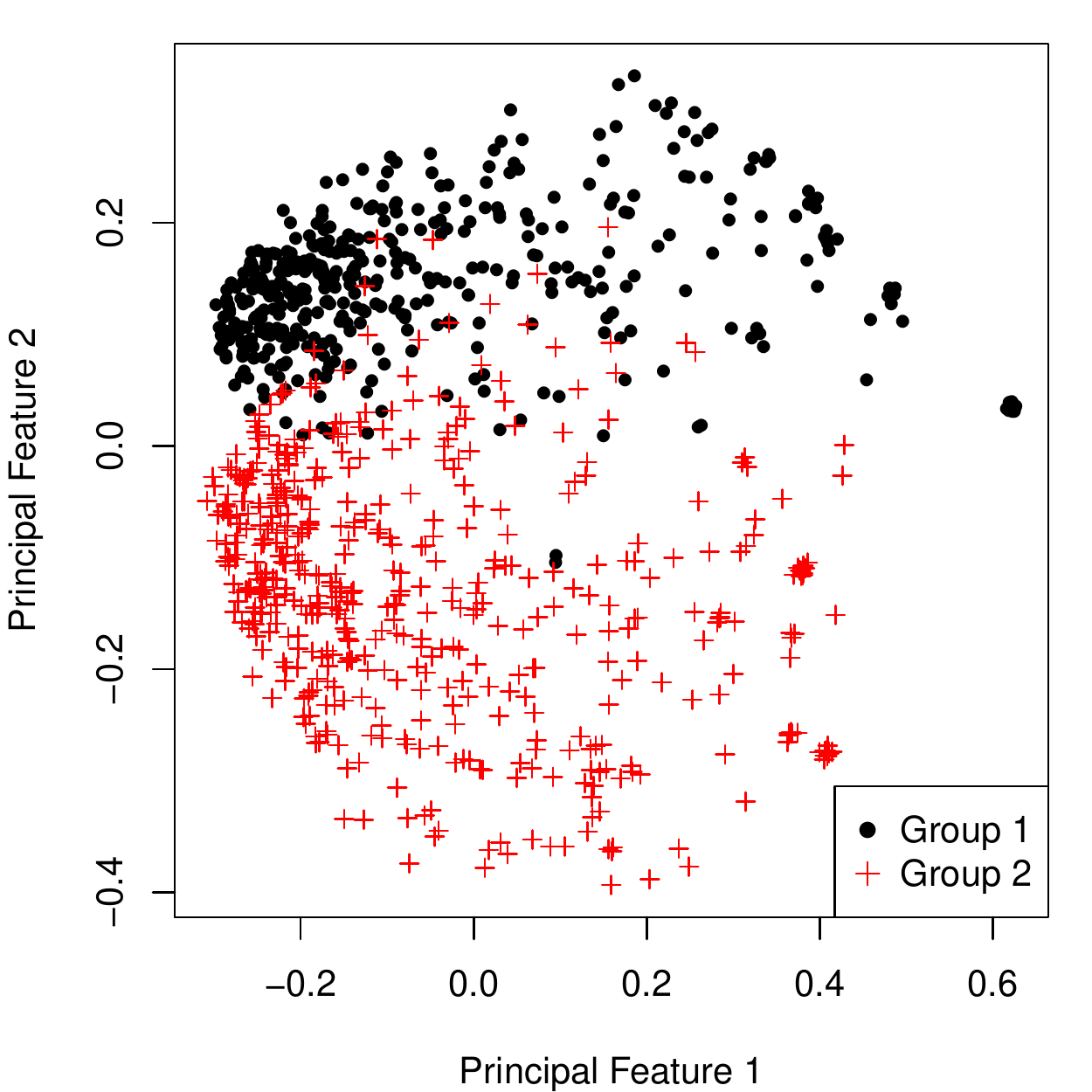}
~
\includegraphics[width=0.45\textwidth]{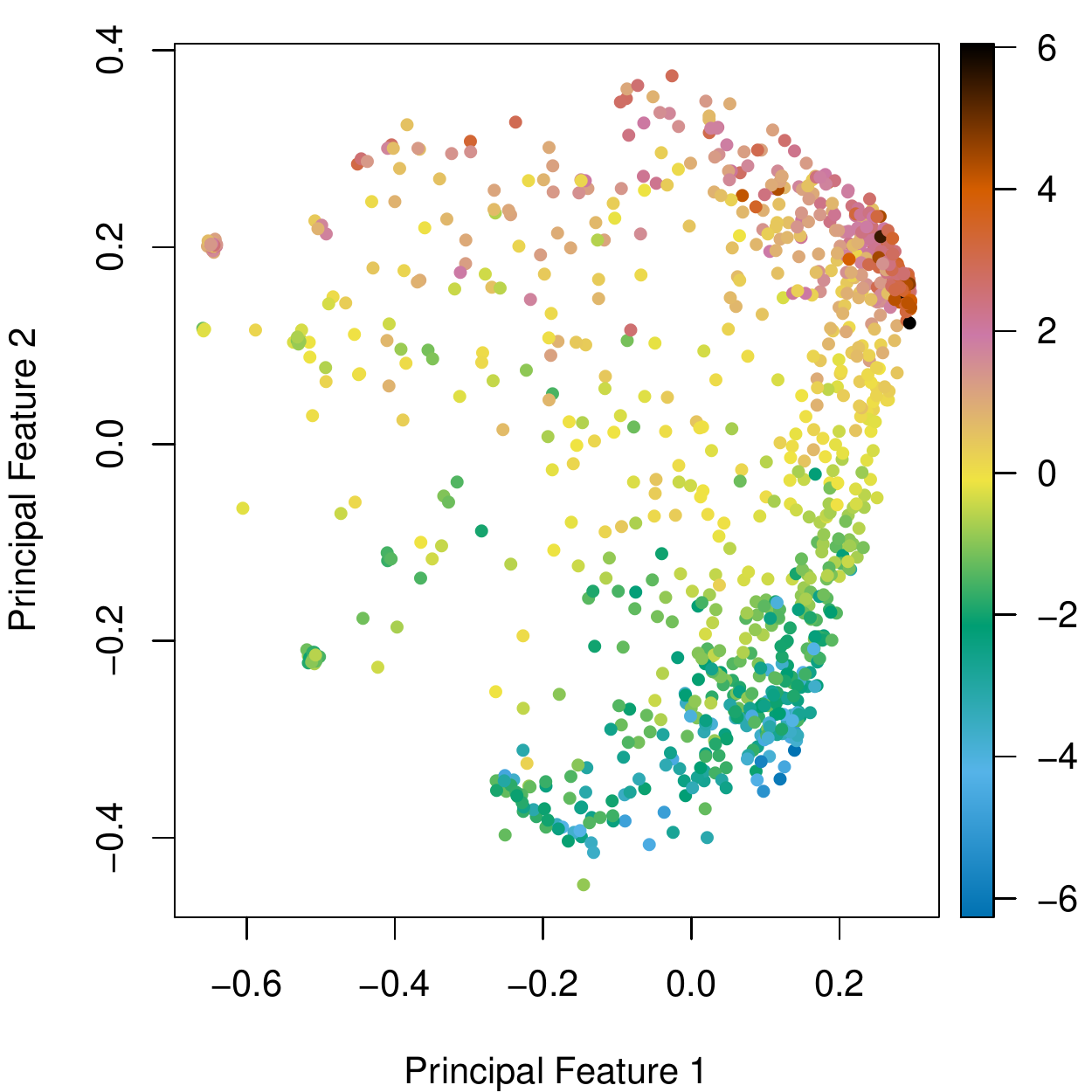}
\caption{First two principal features for one dataset with 1000 sequences generated from strategy I (left) or strategy II (right). The data points in the right panel are colored according to the value of the latent variable $\theta_0$.}\label{fig:sim_latent}
\end{figure}

To take a closer look at how the extracted features reveal the latent structure of action sequences, in Figure \ref{fig:sim_latent}, we plot the first two principal features for one dataset of 1000 sequences under each strategy. For the dataset generated from strategy I (left panel in Figure \ref{fig:sim_latent}), the group structure is clearly shown in the figure and the two groups can be roughly separated by a horizontal line at 0. The data shown in the right panel of Figure \ref{fig:sim_latent} is generated from strategy II. It is evident that sequences located closer have similar latent characteristics.


\section{Case Study}\label{sec:example}
\subsection{Data}
The data considered in this study comes from the PIAAC 2012 survey from five countries: the United Kingdom, Ireland, Japan, the Netherlands and the United States. There are 14 PSTRE items and 11,464 respondents in the dataset in total. Each person responded to all or a subset of the 14 items. There are 7,620 respondents who answered 7 items and 3,645 respondents who answered all 14 items. For each item, there were around 7,500 respondents. Altogether there are 106,096 respondent-item pairs. Both the response process and the response outcome (correct or incorrect) were recorded for each pair. 



Table \ref{table:items} summarizes some basic descriptive statistics of the dataset by item, where $n$ denotes the number of respondents, $N$ is the number of possible actions, $\bar L$ stands for the average process length, and Correct \% is the percentage of correct responses. The 14 items vary in content, task complexity, and difficulty. Items U02 and U04a are the most difficult items as only around 10\% of respondents had the correct answer. The tasks of these two items are also relatively complicated, requiring more than 40 actions on average and having a large number of possible actions. U06a is the simplest item in terms of task complexity since respondents took only 10.8 actions on average to finish the task and the item has the fewest possible actions. Despite the simplicity, less than 30\% of respondents answered U06a correctly. The variety of items necessitates automatic methods to extract features from process data and to avoid identifying important actions and patterns manually, which is time consuming and requires extra work if coding is changed.

\begin{table}[htb]
\begin{threeparttable}
\caption{Descriptive statistics of 14 PIAAC problem-solving items.}\label{table:items}
\centering
\begin{tabular}{llccccc}
\hline
ID & Description & $n$ & $N$ & {\small $\bar L$} &Correct \%\\
\hline
U01a & Party Invitations - Can/Cannot Come & 7620 & 207 & 24.8 & 54.5\\
U01b & Party Invitations - Accommodations & 7670 & 249 & 52.9 & 49.3\\
U02 & Meeting Rooms & 7537 & 328 & 54.1 & 12.8\\
U03a & CD Tally & 7613 & 280 & 13.7 & 37.9\\
U04a & Class Attendance & 7617& 986 & 44.3 & 11.9\\
U06a & Sprained Ankle - Site Evaluation Table & 7622 & 47 & 10.8 & 26.4 \\
U06b & Sprained Ankle - Reliable/Trustworthy Site & 7612 & 98 & 16.0 & 52.3\\
U07 & Digital Photography Book Purchase & 7549 & 125 & 18.6 & 46.0\\
U11b & Locate E-mail - File 3 E-mails & 7528 & 236 & 30.9 & 20.1 \\
U16 & Reply All & 7531 & 257 & 96.9 & 57.0\\
U19a & Club Membership - Member ID & 7556 & 373 & 26.9 & 69.4\\
U19b & Club Membership - Eligibility for Club President & 7558 & 458 & 21.3 & 46.3\\
U21 & Tickets & 7606 & 252 & 23.4 & 38.2\\
U23 & Lamp Return & 7540 & 303 & 28.6 & 34.3\\
\hline
\end{tabular}
\begin{tablenotes}
\item Note: $n$ = number of respondents; $N$ = number of possible actions; $\bar L =$ average process length; Correct \% = percentage of correct responses
\end{tablenotes}
\end{threeparttable}
\end{table} 


\subsection{Feature Interpretation}
We extracted features for each of the 14 items by Procedure \ref{proc:mds_feature}. The number of features is chosen from $\{10, 20, \ldots, 100\}$ by five-fold cross-validation and the selected number for each item is given in the second column of Table \ref{table:feature}.

Many of the principal features, especially the first several ones, have clear interpretations. We find the interpretation of a feature by examining the characteristics of the action sequences corresponding to the two extremes of the feature and then confirm it by calculating the correlation between the feature and a variable constructed according to the interpretation. Table \ref{table:feature} lists the interpretation of the first three principal features for each item.

The first principal feature of each item usually indicates attentiveness. An inattentive respondent often tries to skip a task directly or submits an answer by guessing randomly without meaningful interactions with the simulated environment, while an attentive respondent usually tries to understand and to complete the task by exploring the environment, thus taking more actions. Attentiveness in response process can be reflected in the process length. In Table \ref{table:feature}, the numbers in the parentheses after the interpretation of the first principal feature of each item give the absolute value of the correlation between the first principal feature and the logarithm of the process length. For 13 out of 14 items, the absolute correlation is higher than 0.85. To explore the relation between the 14 first principal features, we multiply the features by the sign of their correlation with the corresponding process length. With the redirection, a higher first principal feature indicates a more attentive respondent. For a given pair of items, we calculate the correlation between their first principal features among the respondents who responded to both items. These correlations range from 0.36 to 0.74, implying that the respondents who tend to skip one item are likely to skip other items as well.

Some other features reveal whether the respondent understands the requirements of items. For example, item U11b requires respondents to classify emails in the ``Save'' folder. The second feature of U11b reflects if a respondent was working on the correct folder. Similarly, item U01b requires creating a new folder. The second feature of this item is related to whether this requirement is followed.

There are also features related to respondents' information and computer technology skills. Examples include the second feature of U03a, indicating whether search or sort tools are used, and the second feature of U04a, reflecting whether window split is used to avoid frequent switching between windows.

\begin{table}
\centering
{\small
\begin{threeparttable}
\caption{Interpretation of first three principal features.}\label{table:feature}
\begin{tabular}{lccc}
\hline
Item & $K$ & Feature & Interpretation \\
\hline
\multirow{3}{*}{U01a} & \multirow{3}{*}{50} & 1 & Attentiveness in item response process (0.68)\\
 & & 2 & Intensity of mail and folder viewing actions\\
 & & 3 & Intensity of mail moving actions\\
\hline
\multirow{3}{*}{U01b} & \multirow{3}{*}{30}& 1 & Attentiveness in item response process (0.96)\\
 & & 2 & Intensity of creating new folders actions\\
 & & 3 & Intensity of mail moving actions\\
\hline
\multirow{3}{*}{U02} & \multirow{3}{*}{50} & 1 & Attentiveness in item response process (0.94)\\
 & & 2 & Intensity of mail moving actions\\
 & & 3 & Intensity of mail viewing actions\\
\hline
\multirow{3}{*}{U03a} & \multirow{3}{*}{70} & 1 & Attentiveness in item response process (0.86)\\
 & & 2 & Intensity of search and sort actions\\
 & & 3 & Times of answer submission\\
\hline
\multirow{3}{*}{U04a} & \multirow{3}{*}{70} & 1 & Attentiveness in item response process (0.98)\\
 & & 2 & Intensity of switching environments\\
 & & 3 & Intensity of arranging tables actions\\
\hline
\multirow{3}{*}{U06a} & \multirow{3}{*}{60} & 1 & Attentiveness in item response process (0.91)\\
 & & 2 & Intensity of clicking radio buttons\\
 & & 3 & Chance of classifying a website as useful\\
\hline
\multirow{3}{*}{U06b} & \multirow{3}{*}{20} & 1 & Attentiveness in item response process (0.94)\\
 & & 2 & Intensity of selecting answers\\
 & & 3 & Intensity of choosing website 2 against choosing website 4\\
\hline
\multirow{3}{*}{U07} & \multirow{3}{*}{100} & 1 & Attentiveness in item response process (0.96)\\
 & & 2 & Intensity of actions related to website 6\\
 & & 3 & Intensity of actions related to website 3\\
\hline
\multirow{3}{*}{U11b} & \multirow{3}{*}{40} & 1 & Attentiveness in item response process (0.94)\\
 & & 2 & Intensity of actions related to email in save folder\\
 & & 3 & Intensity of mail moving actions\\
\hline
\multirow{3}{*}{U16} & \multirow{3}{*}{70} & 1 & Attentiveness in item response process (0.95)\\
 & & 2 & Intensity of ``Other\_Keypress''\\
 & & 3 & Intensity of email viewing against email replying\\
\hline
\multirow{3}{*}{U19a} & \multirow{3}{*}{40} & 1 & Attentiveness in item response process (0.91)\\
 & & 2 & Intensity of typing emails\\
 & & 3 & Intensity of ticking and clicking email environment button\\
\hline
\multirow{3}{*}{U19b} & \multirow{3}{*}{50} & 1 & Attentiveness in item response process (0.89)\\
 & & 2 & Intensity of sorting actions\\
 & & 3 & Number of checked boxes\\
\hline
\multirow{3}{*}{U21} & \multirow{3}{*}{50} & 1 & Attentiveness in item response process (0.92)\\
 & & 2 & Intensity of making reservations\\
 & & 3 & Number of games selected\\
\hline
\multirow{3}{*}{U23} & \multirow{3}{*}{40} & 1 & Attentiveness in item response process (0.87)\\
 & & 2 & Click customer service against clicking not needed links \\
 & & 3 & Obtain Authorization number or not\\
\hline
\end{tabular}
\begin{tablenotes}
\item Note: Number in parentheses represents absolute value of correlation between first principal feature and logarithm of sequence length.
\end{tablenotes}
\end{threeparttable}}
\end{table}

\subsection{Reconstruction of Derived Variables}
In this subsection, we demonstrate that the extracted features contain a substantial amount of information of the action sequences by showing that some key variables derived from the action sequences can be reconstructed from the features.

Derived variables are binary variables indicating whether certain actions or patterns appear in the action sequences. For the example item described in the introduction, whether the first link is clicked is a derived variable. Item response outcomes (correct or incorrect) can also be treated as derived variables since they are entirely determined by the action sequences. In PIAAC data, besides the item response outcomes, 79 derived variables are recorded for the 14 items. The following experiment examines how well the 93 (79 + 14) derived variables can be reconstructed from the features extracted from Procedure \ref{proc:mds_feature}.

For a given item, let $Y$ denote a generic binary derived variable and $\bm x$ be a vector of principal features extracted from its response process. We consider the logistic regression model for each derived variable
\begin{equation}\label{eq:model_derived_var}
\log \left(\frac{p}{1-p}\right)= \bm \eta^T \bm \beta,
\end{equation}
where $p$ is the probability of $Y = 1$ and $\bm \eta^T = (1, \bm x^T)$. 
For each derived variable, the respondents with the variable are randomly divided into a training set and test set in the ratio 4:1. The logistic regression model \eqref{eq:model_derived_var} is fit on the training set and the value of derived variable in the test set is predicted as 1 if the fitted probability is greater than 0.5, and 0 otherwise. The prediction performance is evaluated by prediction accuracy.

\begin{figure}[htb]
\centering
\includegraphics[width=10cm]{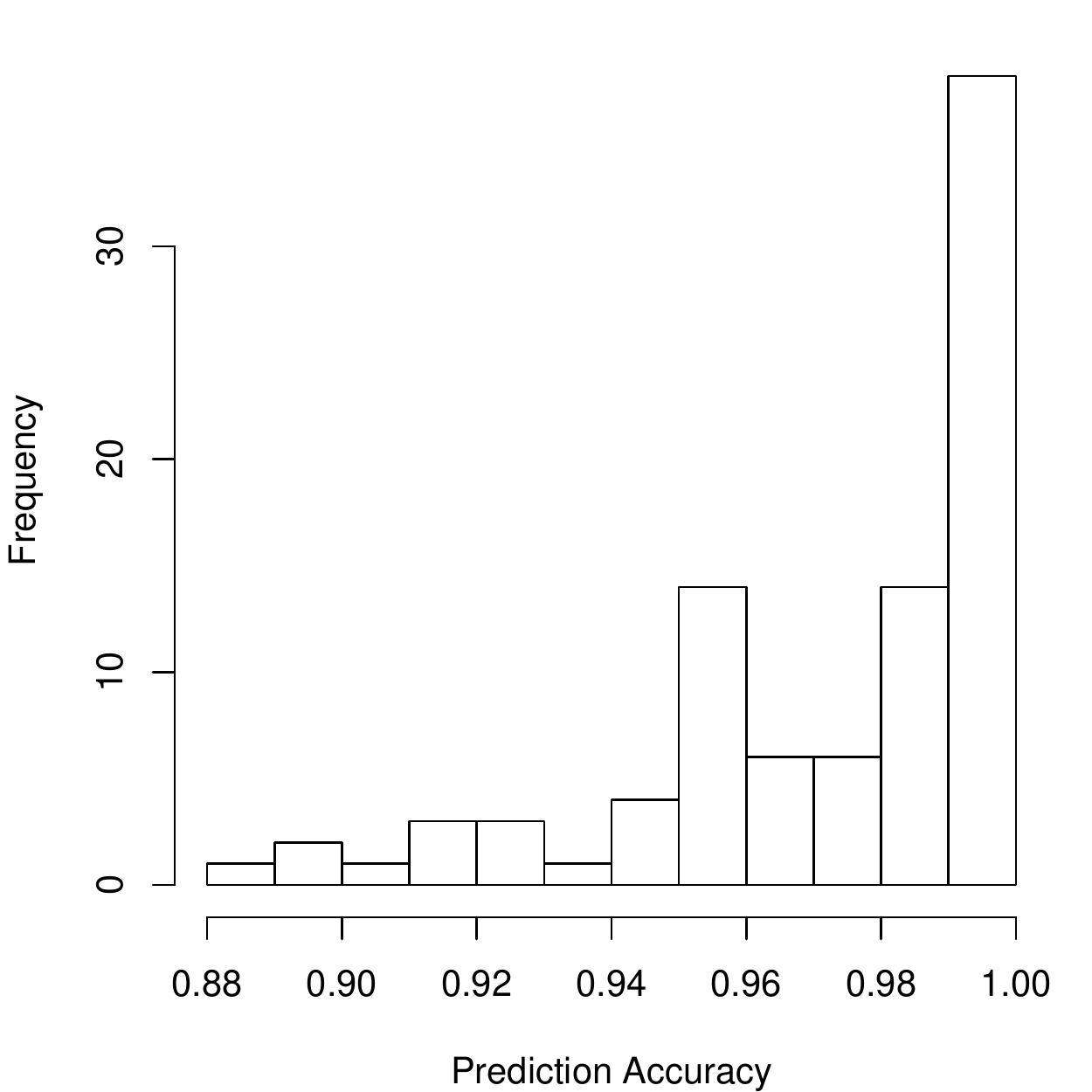}
\caption{Histogram of the prediction accuracy of derived variables.}\label{fig:hist_acc_derived_var}
\end{figure}

Figure \ref{fig:hist_acc_derived_var} presents a histogram of the prediction accuracy for the 93 derived variables. 
For most of the variables, the model constructed from the extracted features has more than 90\% accuracy.
This result confirms that the features extracted by Procedure \ref{proc:mds_feature} is a comprehensive summary of the response processes.

Given that the features contain information about action sequences, a natural question is whether these features are useful for assessing respondents' competency and understanding their behavior. We will try to answer this question in the remainder of this section.


\subsection{Cross-Item Outcome Prediction}
In this section, we explore if the features obtained from the process data of one item are helpful to predict the outcomes of another item. Intuitively, if the extracted features characterize the behavioral patterns and/or intellectual levels of respondents, which affect their performance in general, then these features should be able to tell more about whether the respondents can answer other items correctly than a single binary outcome. 

Let $Y_j$ denote the outcome of item $j$ and $\bm x_j \in \mathbb{R}^{K_j}$ denote the features extracted from item $j$, $j=1, \ldots, 14$. We model the relation between the outcome of item $j$ and the outcome and the features of item $j' \neq j$ by a logistic regression
\begin{equation}\label{eq:outcome_pred_model}
\log\left(\frac{p_j}{1- p_j}\right)= \bm \eta_{j'}^T \bm \beta,
\end{equation}
where $p_j$ is the probability of $Y_j = 1$ and $\bm \eta_{j'}$ is a vector of covariates of item $j'$. If process data is not taken into account, only $Y_{j'}$ provides information about $Y_j$ and $\bm \eta_{j'}^T = (1, Y_{j'})$. In this case, available information for telling the outcome of item $j$ is very limited, especially when the correct rate of item $j'$ is close to 0 or 1. If process data is collected, then the features extracted according to Procedure \ref{proc:mds_feature} provide another source of information and we could use $\bm \eta^T = (1, Y_{j'}, \bm x_{j'}^T, Y_{j'} \bm x_{j'}^T)$ as the covariates from item $j'$. We call it the baseline model if it only incorporates the outcome in $\bm \eta$ and the process model if it utilizes the features extracted from process data.

\begin{figure}[htb]
\includegraphics[width=\textwidth]{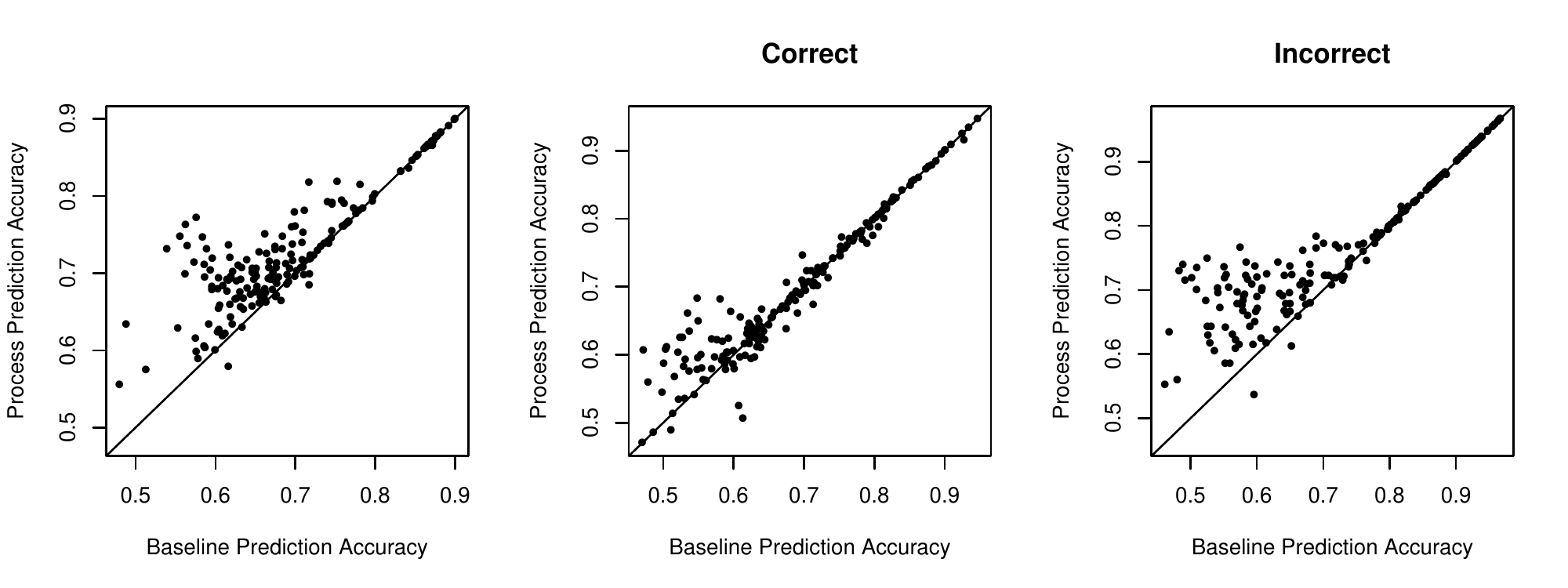}
\caption{Left: Prediction accuracy of the process model against the baseline model; Middle: Prediction accuracy of the process model against the baseline model for respondents who answered the predictor item correctly; Right: Prediction accuracy of the process model against the baseline model for respondents who answered the predictor item incorrectly. }\label{fig:outcome_pred}
\end{figure}

Given that we want to model the outcome of item $j$ based on the information provided in item $j'$, respondents who responded to both items are randomly split into training, validation, and test sets in the ratio 4:1:1. Both the baseline model and the process model are fit on the training set. To avoid overfitting in the process model, $L_2$ penalties on the coefficients are incorporated. The process model is fitted on the training set for a grid of penalty parameters. The fitted process model that corresponds to the penalty parameter producing the highest prediction accuracy on the validation set is chosen to compare with the baseline model. 
The prediction accuracy of the process model for all combinations of $j$ and $j'$ is plotted against the prediction accuracy of the corresponding baseline model in the left panel of Figure \ref{fig:outcome_pred}. For most of the item pairs, the prediction accuracy is improved when the features extracted from process data are utilized, implying that the information in the process data is helpful in predicting the performance of respondents.

\begin{figure}[htb]
\centering
\includegraphics[width=\textwidth]{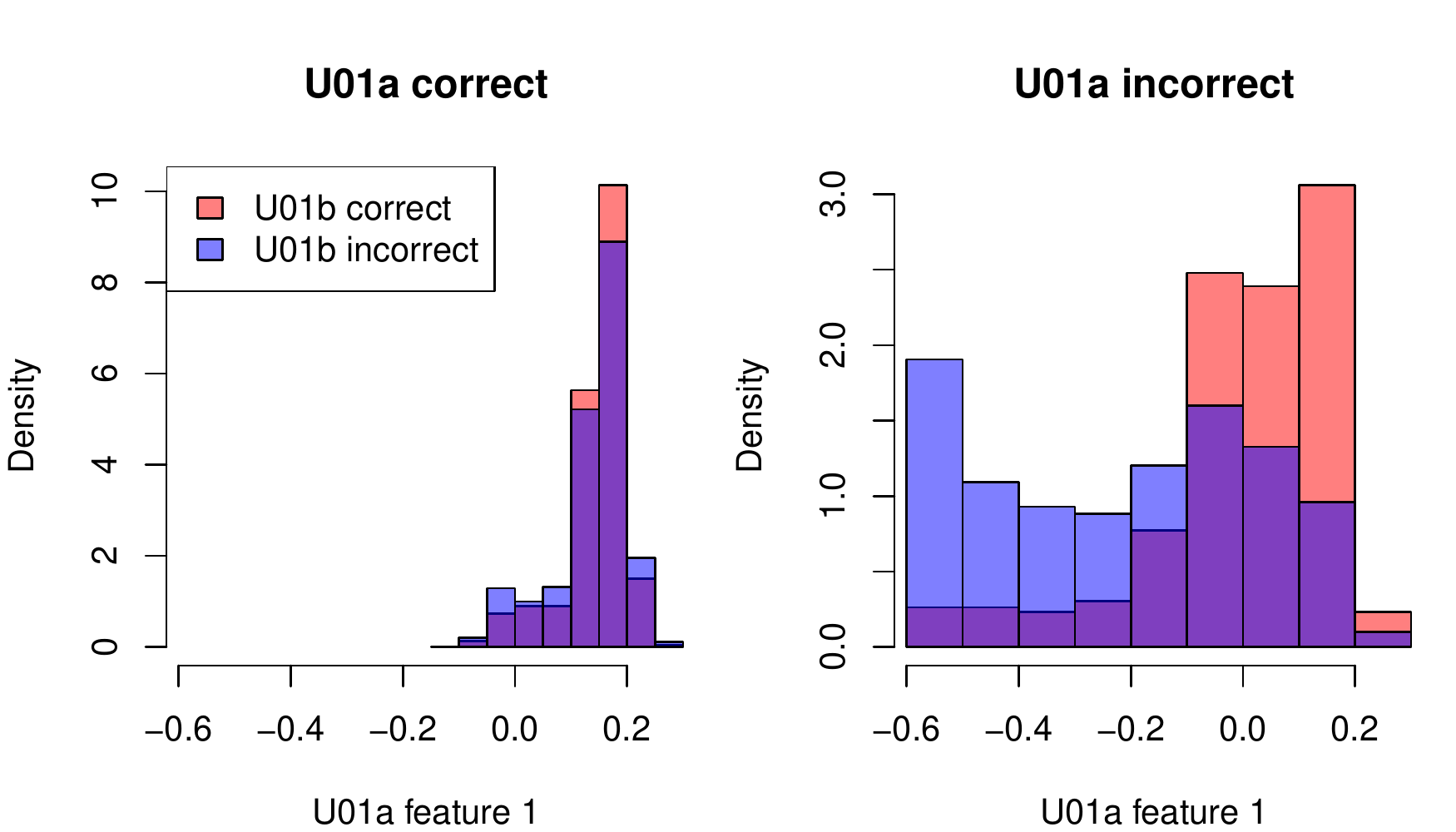}
\caption{Histograms of the first principal feature of U01a stratified by the outcomes of U01a and U01b.}\label{fig:hist_feature}
\end{figure}

To take a closer look at the results, the middle and right panels of Figure \ref{fig:outcome_pred} compare prediction accuracy separately for those who answered item $j'$ correctly and incorrectly. The improvement in prediction accuracy is more obvious for the ``incorrect'' group. The main reason is that the action sequences corresponding to the incorrect responses usually provide more information about the respondents. There are usually more ways to answer a question incorrectly than correctly. An incorrect response may be the consequence of misunderstanding the item requirements or lack of basic computer skills. It may also result from the respondents' carelessness or inattentiveness. These varieties are reflected in the response processes, and thus, in the extracted features. As an illustration, the histograms of the first principal feature of item U01a stratified by the respondents' outcomes of U01a and U01b are plotted in Figure \ref{fig:hist_feature}. 
In the U01a incorrect group, there is a significant difference in the feature distributions for those who answered U01b correctly and incorrectly, while the two distributions are almost identical in the U01a correct group. Recall that the first principal feature describes the respondents' attentiveness. Among the respondents who answer U01a incorrectly, those with lower feature values lack attentiveness. By including the features in the model, we are able to identify them and know that they are unlikely to answer U01b and other items correctly. 

\subsection{Score Prediction}
The 14 interactive items in PIAAC were designed to study the PSTRE skills. The respondents' competency in literacy and numeracy were measured using items designed specifically for these two scales. We will show in this subsection that the process data from problem-solving items can cast light on respondents' proficiency in other scales.
Let $Z$ denote the score of a specific scale. We consider a linear model to explore the relation between $Z$ and problem-solving items 
\begin{equation}\label{eq:score_linear_model}
Z = \bm \eta ^T \bm \beta + \varepsilon,
\end{equation} 
where $\varepsilon$ is a Gaussian random noise and $\bm \eta$ is a vector of predictors related to one or more problem-solving items and will be specified later.

\subsubsection{Score Prediction Using a Single Item}
In the first experiment, we model the scores based on the information provided in a single item. 
In the model that only incorporates the binary outcome, namely the baseline model, the linear predictor is $\bm \eta^T = (1, Y_j)$. In the process model, we use $\bm \eta^T = (1, Y_j, \bm {x}_j^{T},  Y_j\bm {x}_{j}^T)$. For each of the 14 problem-solving items, the respondents are randomly split into training, validation and test sets in the ratio 4:1:1. Both the baseline and the process model are fitted on the training set for literacy and numeracy scores separately. To avoid overfitting, $L_2$ penalties are placed on the coefficients in the process model for a grid of penalty parameters. The penalty parameter that produces the best prediction performance on the validation set is selected to obtain the final estimated process model. The prediction performance is evaluated by $\text{OSR}^2$.

\begin{figure}
\centering
\includegraphics[width=0.45\textwidth]{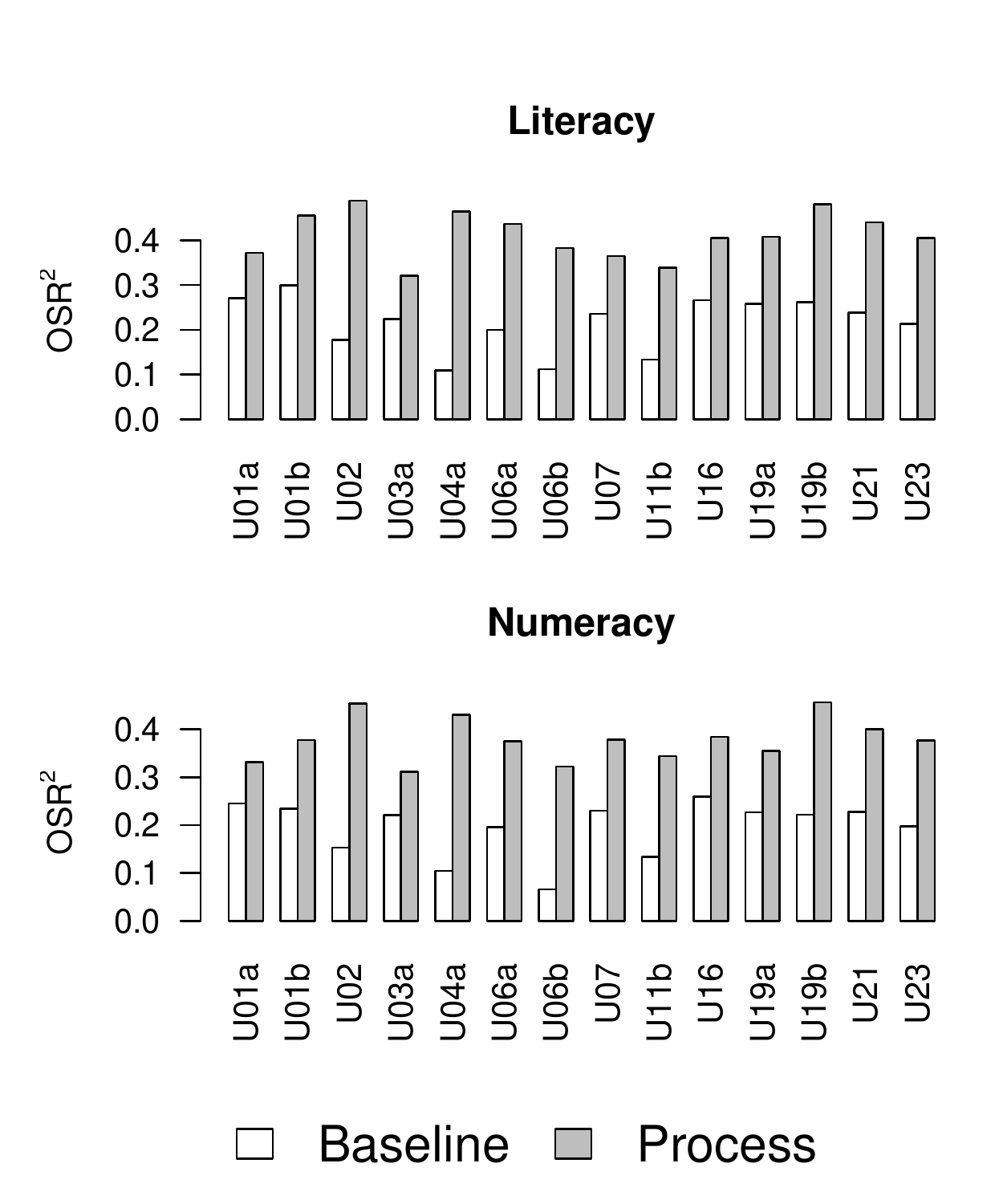}
\includegraphics[width=0.45\textwidth]{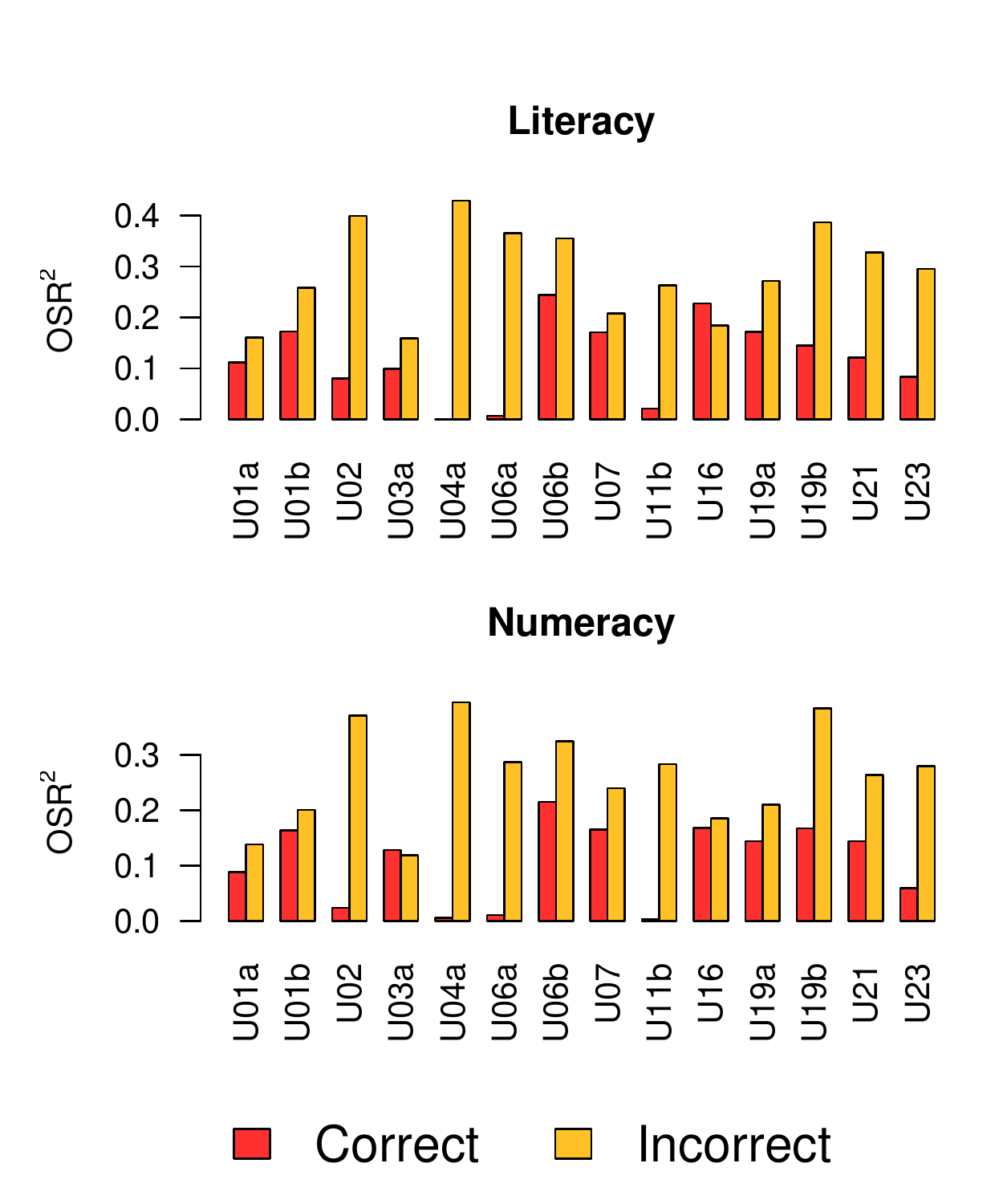}
\caption{Left: OSR${}^2$ of the baseline and process model on the test set. Right: OSR${}^2$ of the process model stratified by outcomes.}\label{fig:score_pred}
\end{figure}

The left panel of Figure \ref{fig:score_pred} presents the $\text{OSR}^2$ of the baseline model and the process model for all combinations of score and item. For both literacy and numeracy scores, including information from process data is beneficial to score prediction. Although the problem-solving items are not designed to measure numeracy and literacy in PIAAC, process data can provide information leading to substantial improvements in these two scales.

The right panel of Figure \ref{fig:score_pred} presents $\text{OSR}^2$ of the process model stratified by the outcome of an item. Similar to the outcome prediction in the previous subsection, the prediction performance for the respondents who answered an item incorrectly is usually much better than that for those who answered correctly since action sequences corresponding to incorrect answers often have more information than those corresponding to correct answers.

\subsubsection{Score Prediction Using Multiple Items}
In the second experiment, we will examine how the improvement in score prediction brought by process data changes as the number of available items increases. 
We only consider the 3,645 respondents who responded to all 14 problem-solving items in this experiment. Among these respondents, 2,645 are randomly assigned to the training set, 500 to the validation set and 500 to the test set. 
For each score, two models, a baseline model and a process model, are considered for a given set of available items. For the baseline model, the linear predictor consists of the binary outcomes of the available items. For the process model, in addition to the binary outcomes, the linear predictor includes the first 20 principal features for each available item. Let $S_m = \{j_1, \ldots, j_m\}$ be the indices of the available items. Then the linear predictor for the baseline model is $\bm \eta^T = (1, Y_{j_1}, \ldots, Y_{j_m})$, while the linear predictor of the process model is $\bm \eta^T = (1, Y_{j_1}, \ldots, Y_{j_m}, \bm x_{j_1}, \ldots, \bm x_{j_m})$ where $\bm x_{j} \in \mathbb{R}^{20}$ is the first 20 principal features for item $j$. The set of available items is determined by forward Akaike information criterion (AIC) selection of the outcomes on the training set. Specifically, for a given $m$, $S_m$ contains the items whose outcomes are the first $m$ outcomes selected by the forward AIC selection among all 14 outcomes $Y_1, \ldots, Y_{14}$. For a given score, a sequence of baseline models and the process models are fitted on the training set. Similar to the previous subsection, $L_2$ penalty is added on the coefficients of the process models to avoid overfitting, and the penalty parameter is selected based on the $\text{OSR}^2$ on the validation set.

\begin{figure}
\centering
\includegraphics[width=\textwidth]{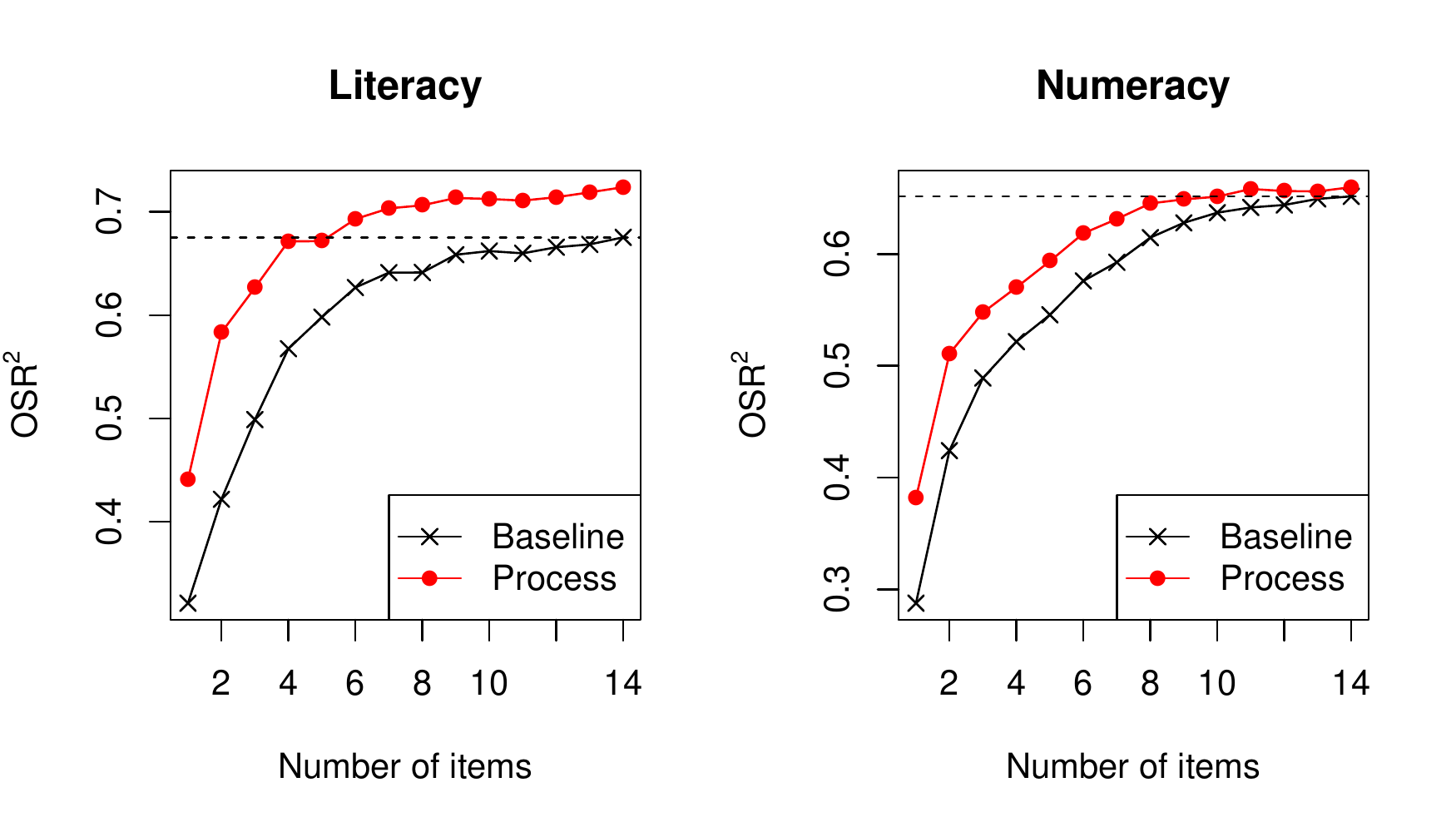}
\caption{OSR${}^{2}$ of the baseline and process model with various number of items.}\label{fig:score_pred_all}
\end{figure}

Figure \ref{fig:score_pred_all} presents the $\text{OSR}^2$ of the baseline model and the selected process model on the test set.
Regardless of the number of items available, the process model outperforms the baseline model in both literacy and numeracy score prediction. The improvement is more significant for literacy. The $\text{OSR}^2$ of the process model with only five items is comparable to the $\text{OSR}^2$ of the baseline model with all 14 items. In the process of completing the task in the problem-solving item, respondents need to comprehend the item description and provided materials, so the outcomes and the action sequences of problem-solving items can reflect respondents' literacy competency to some extent. Our experiment shows that process data can provide more information that binary outcomes. Properly incorporating process data in data analysis can exploit the information from items more efficiently.

\section{Concluding Remarks}\label{sec:discussion}
In this article, we present a method to extract informative latent variables from process data and illustrate the method via simulation studies and a case study of PIAAC 2012 data. 
The latent variables in the process data are extracted by an automatic procedure involving MDS of the dissimilarity matrix among response processes.
The dissimilarity measure used in this article is just one of the possible choices. Other choices such as Levenshtein distance \citep{levenshtein1966binary} and optimal symbol alignment distance \citep{herranz2011optimal} can also be used and give similar results. However, these measures are often more computationally demanding.

The respondents of our process data came from five different countries and they varied in age, gender, and many other demographic variables. The extracted features and the prediction procedure can also be used to study the difference in behavior patterns of different demographic groups. We will pursue this direction in future research.

Time stamps of actions are also available in process data. The time elapsed between the occurrences of two actions may provide additional information about respondents and can be useful in cognitive assessments. The current dissimilarity measure does not make use of this information. Further study on incorporating response time information in the analysis of process data is a potential future direction.

\section{Acknowledgments}
The authors would like to thank Educational Testing Service for providing the data, and Hok Kan Ling for cleaning it.

\bibliography{mds}

\begin{thebibliography}{}

\bibitem [\protect \citeauthoryear {%
Borg%
\ \BBA {} Groenen%
}{%
Borg%
\ \BBA {} Groenen%
}{%
{\protect \APACyear {2005}}%
}]{%
borg2005modern}
\APACinsertmetastar {%
borg2005modern}%
\begin{APACrefauthors}%
Borg, I.%
\BCBT {}\ \BBA {} Groenen, P\BPBI J.%
\end{APACrefauthors}%
\unskip\
\newblock
\APACrefYear{2005}.
\newblock
\APACrefbtitle {Modern multidimensional scaling: Theory and applications}
  {Modern multidimensional scaling: Theory and applications}.
\newblock
\APACaddressPublisher{New York, NY}{Springer Science \& Business Media}.
\newblock
\begin{APACrefDOI} \doi{10.1007/0-387-28981-X} \end{APACrefDOI}
\PrintBackRefs{\CurrentBib}

\bibitem [\protect \citeauthoryear {%
G{\'o}mez-Alonso%
\ \BBA {} Valls%
}{%
G{\'o}mez-Alonso%
\ \BBA {} Valls%
}{%
{\protect \APACyear {2008}}%
}]{%
gomez2008similarity}
\APACinsertmetastar {%
gomez2008similarity}%
\begin{APACrefauthors}%
G{\'o}mez-Alonso, C.%
\BCBT {}\ \BBA {} Valls, A.%
\end{APACrefauthors}%
\unskip\
\newblock
\APACrefYearMonthDay{2008}{}{}.
\newblock
{\BBOQ}\APACrefatitle {A similarity measure for sequences of categorical data
  based on the ordering of common elements} {A similarity measure for sequences
  of categorical data based on the ordering of common elements}.{\BBCQ}
\newblock
\BIn{} V.~Torra\ \BBA {} Y.~Narukawa\ (\BEDS), \APACrefbtitle {Modeling
  Decisions for Artificial Intelligence} {Modeling decisions for artificial
  intelligence}\ (\BPGS\ 134--145).
\newblock
\APACaddressPublisher{Berlin, Heidelberg}{Springer Berlin Heidelberg}.
\newblock
\begin{APACrefDOI} \doi{https://doi.org/10.1007/978-3-540-88269-5_13}
  \end{APACrefDOI}
\PrintBackRefs{\CurrentBib}

\bibitem [\protect \citeauthoryear {%
Greiff%
, Niepel%
, Scherer%
\BCBL {}\ \BBA {} Martin%
}{%
Greiff%
\ \protect \BOthers {.}}{%
{\protect \APACyear {2016}}%
}]{%
greiff2016understanding}
\APACinsertmetastar {%
greiff2016understanding}%
\begin{APACrefauthors}%
Greiff, S.%
, Niepel, C.%
, Scherer, R.%
\BCBL {}\ \BBA {} Martin, R.%
\end{APACrefauthors}%
\unskip\
\newblock
\APACrefYearMonthDay{2016}{}{}.
\newblock
{\BBOQ}\APACrefatitle {Understanding students' performance in a computer-based
  assessment of complex problem solving: An analysis of behavioral data from
  computer-generated log files} {Understanding students' performance in a
  computer-based assessment of complex problem solving: An analysis of
  behavioral data from computer-generated log files}.{\BBCQ}
\newblock
\APACjournalVolNumPages{Computers in Human Behavior}{61}{}{36--46}.
\newblock
\begin{APACrefDOI} \doi{10.1016/j.chb.2016.02.095} \end{APACrefDOI}
\PrintBackRefs{\CurrentBib}

\bibitem [\protect \citeauthoryear {%
He%
\ \BBA {} von Davier%
}{%
He%
\ \BBA {} von Davier%
}{%
{\protect \APACyear {2015}}%
}]{%
he2015identifying}
\APACinsertmetastar {%
he2015identifying}%
\begin{APACrefauthors}%
He, Q.%
\BCBT {}\ \BBA {} von Davier, M.%
\end{APACrefauthors}%
\unskip\
\newblock
\APACrefYearMonthDay{2015}{}{}.
\newblock
{\BBOQ}\APACrefatitle {Identifying feature sequences from process data in
  problem-solving items with n-grams} {Identifying feature sequences from
  process data in problem-solving items with n-grams}.{\BBCQ}
\newblock
\BIn{} L\BPBI A.~van~der Ark, D\BPBI M.~Bolt, W\BHBI C.~Wang, J\BPBI
  A.~Douglas\BCBL {}\ \BBA {} S\BHBI M.~Chow\ (\BEDS), \APACrefbtitle
  {Quantitative Psychology Research} {Quantitative psychology research}\
  (\BPGS\ 173--190).
\newblock
\APACaddressPublisher{Cham}{Springer International Publishing}.
\newblock
\begin{APACrefDOI} \doi{https://doi.org/10.1007/978-3-319-19977-1_13}
  \end{APACrefDOI}
\PrintBackRefs{\CurrentBib}

\bibitem [\protect \citeauthoryear {%
He%
\ \BBA {} von Davier%
}{%
He%
\ \BBA {} von Davier%
}{%
{\protect \APACyear {2016}}%
}]{%
he2016analyzing}
\APACinsertmetastar {%
he2016analyzing}%
\begin{APACrefauthors}%
He, Q.%
\BCBT {}\ \BBA {} von Davier, M.%
\end{APACrefauthors}%
\unskip\
\newblock
\APACrefYearMonthDay{2016}{}{}.
\newblock
{\BBOQ}\APACrefatitle {Analyzing process data from problem-solving items with
  n-grams: Insights from a computer-based large-scale assessment} {Analyzing
  process data from problem-solving items with n-grams: Insights from a
  computer-based large-scale assessment}.{\BBCQ}
\newblock
\BIn{} Y.~Rosen, S.~Ferrara\BCBL {}\ \BBA {} M.~Mosharraf\ (\BEDS),
  \APACrefbtitle {Handbook of research on technology tools for real-world skill
  development} {Handbook of research on technology tools for real-world skill
  development}\ (\BPGS\ 749--776).
\newblock
\APACaddressPublisher{Hershey, PA}{Information Science Reference}.
\newblock
\begin{APACrefDOI} \doi{10.4018/978-1-4666-9441-5.ch029} \end{APACrefDOI}
\PrintBackRefs{\CurrentBib}

\bibitem [\protect \citeauthoryear {%
Herranz%
, Nin%
\BCBL {}\ \BBA {} Sole%
}{%
Herranz%
\ \protect \BOthers {.}}{%
{\protect \APACyear {2011}}%
}]{%
herranz2011optimal}
\APACinsertmetastar {%
herranz2011optimal}%
\begin{APACrefauthors}%
Herranz, J.%
, Nin, J.%
\BCBL {}\ \BBA {} Sole, M.%
\end{APACrefauthors}%
\unskip\
\newblock
\APACrefYearMonthDay{2011}{}{}.
\newblock
{\BBOQ}\APACrefatitle {Optimal symbol alignment distance: A new distance for
  sequences of symbols} {Optimal symbol alignment distance: A new distance for
  sequences of symbols}.{\BBCQ}
\newblock
\APACjournalVolNumPages{IEEE Transactions on Knowledge and Data
  Engineering}{23}{10}{1541--1554}.
\newblock
\begin{APACrefDOI} \doi{10.1109/TKDE.2010.190} \end{APACrefDOI}
\PrintBackRefs{\CurrentBib}

\bibitem [\protect \citeauthoryear {%
Karni%
\ \BBA {} Levin%
}{%
Karni%
\ \BBA {} Levin%
}{%
{\protect \APACyear {1972}}%
}]{%
karni1972use}
\APACinsertmetastar {%
karni1972use}%
\begin{APACrefauthors}%
Karni, E\BPBI S.%
\BCBT {}\ \BBA {} Levin, J.%
\end{APACrefauthors}%
\unskip\
\newblock
\APACrefYearMonthDay{1972}{}{}.
\newblock
{\BBOQ}\APACrefatitle {The use of smallest space analysis in studying scale
  structure: An application to the California Psychological Inventory.} {The
  use of smallest space analysis in studying scale structure: An application to
  the california psychological inventory.}{\BBCQ}
\newblock
\APACjournalVolNumPages{Journal of Applied Psychology}{56}{4}{341}.
\newblock
\begin{APACrefDOI} \doi{10.1037/h0032934} \end{APACrefDOI}
\PrintBackRefs{\CurrentBib}

\bibitem [\protect \citeauthoryear {%
Klein~Entink%
, Fox%
\BCBL {}\ \BBA {} van~der Linden%
}{%
Klein~Entink%
\ \protect \BOthers {.}}{%
{\protect \APACyear {2009}}%
}]{%
entink2009multivariate}
\APACinsertmetastar {%
entink2009multivariate}%
\begin{APACrefauthors}%
Klein~Entink, R.%
, Fox, J\BHBI P.%
\BCBL {}\ \BBA {} van~der Linden, W\BPBI J.%
\end{APACrefauthors}%
\unskip\
\newblock
\APACrefYearMonthDay{2009}{}{}.
\newblock
{\BBOQ}\APACrefatitle {A multivariate multilevel approach to the modeling of
  accuracy and speed of test takers} {A multivariate multilevel approach to the
  modeling of accuracy and speed of test takers}.{\BBCQ}
\newblock
\APACjournalVolNumPages{Psychometrika}{74}{1}{21}.
\newblock
\begin{APACrefDOI} \doi{10.1007/s11336-008-9075-y} \end{APACrefDOI}
\PrintBackRefs{\CurrentBib}

\bibitem [\protect \citeauthoryear {%
Kroehne%
\ \BBA {} Goldhammer%
}{%
Kroehne%
\ \BBA {} Goldhammer%
}{%
{\protect \APACyear {2018}}%
}]{%
kroehne2018conceptualize}
\APACinsertmetastar {%
kroehne2018conceptualize}%
\begin{APACrefauthors}%
Kroehne, U.%
\BCBT {}\ \BBA {} Goldhammer, F.%
\end{APACrefauthors}%
\unskip\
\newblock
\APACrefYearMonthDay{2018}{}{}.
\newblock
{\BBOQ}\APACrefatitle {How to conceptualize, represent, and analyze log data
  from technology-based assessments? {A} generic framework and an application
  to questionnaire items} {How to conceptualize, represent, and analyze log
  data from technology-based assessments? {A} generic framework and an
  application to questionnaire items}.{\BBCQ}
\newblock
\APACjournalVolNumPages{Behaviormetrika}{45}{2}{527--563}.
\newblock
\begin{APACrefDOI} \doi{https://doi.org/10.1007/s41237-018-0063-y}
  \end{APACrefDOI}
\PrintBackRefs{\CurrentBib}

\bibitem [\protect \citeauthoryear {%
Levenshtein%
}{%
Levenshtein%
}{%
{\protect \APACyear {1966}}%
}]{%
levenshtein1966binary}
\APACinsertmetastar {%
levenshtein1966binary}%
\begin{APACrefauthors}%
Levenshtein, V\BPBI I.%
\end{APACrefauthors}%
\unskip\
\newblock
\APACrefYearMonthDay{1966}{}{}.
\newblock
{\BBOQ}\APACrefatitle {Binary codes capable of correcting deletions,
  insertions, and reversals} {Binary codes capable of correcting deletions,
  insertions, and reversals}.{\BBCQ}
\newblock
\APACjournalVolNumPages{Soviet Physics Doklady}{10}{8}{707--710}.
\PrintBackRefs{\CurrentBib}

\bibitem [\protect \citeauthoryear {%
Lord%
}{%
Lord%
}{%
{\protect \APACyear {1980}}%
}]{%
lord1980applications}
\APACinsertmetastar {%
lord1980applications}%
\begin{APACrefauthors}%
Lord, F\BPBI M.%
\end{APACrefauthors}%
\unskip\
\newblock
\APACrefYear{1980}.
\newblock
\APACrefbtitle {Applications of Item Response Theory to Practical Testing
  Problems} {Applications of item response theory to practical testing
  problems}.
\newblock
\APACaddressPublisher{New York, NY}{Routledge}.
\PrintBackRefs{\CurrentBib}

\bibitem [\protect \citeauthoryear {%
Meyer%
\ \BBA {} Reynolds%
}{%
Meyer%
\ \BBA {} Reynolds%
}{%
{\protect \APACyear {2018}}%
}]{%
meyer2018scores}
\APACinsertmetastar {%
meyer2018scores}%
\begin{APACrefauthors}%
Meyer, E\BPBI M.%
\BCBT {}\ \BBA {} Reynolds, M\BPBI R.%
\end{APACrefauthors}%
\unskip\
\newblock
\APACrefYearMonthDay{2018}{}{}.
\newblock
{\BBOQ}\APACrefatitle {Scores in Space: Multidimensional Scaling of the WISC-V}
  {Scores in space: Multidimensional scaling of the wisc-v}.{\BBCQ}
\newblock
\APACjournalVolNumPages{Journal of Psychoeducational
  Assessment}{36}{6}{562-575}.
\newblock
\begin{APACrefDOI} \doi{10.1177/0734282917696935} \end{APACrefDOI}
\PrintBackRefs{\CurrentBib}

\bibitem [\protect \citeauthoryear {%
Qian%
, Staniewska%
, Reckase%
\BCBL {}\ \BBA {} Woo%
}{%
Qian%
\ \protect \BOthers {.}}{%
{\protect \APACyear {2016}}%
}]{%
qian2016using}
\APACinsertmetastar {%
qian2016using}%
\begin{APACrefauthors}%
Qian, H.%
, Staniewska, D.%
, Reckase, M.%
\BCBL {}\ \BBA {} Woo, A.%
\end{APACrefauthors}%
\unskip\
\newblock
\APACrefYearMonthDay{2016}{}{}.
\newblock
{\BBOQ}\APACrefatitle {Using Response Time to Detect Item Preknowledge in
  Computer-Based Licensure Examinations} {Using response time to detect item
  preknowledge in computer-based licensure examinations}.{\BBCQ}
\newblock
\APACjournalVolNumPages{Educational Measurement: Issues and
  Practice}{35}{1}{38--47}.
\newblock
\begin{APACrefDOI} \doi{10.1111/emip.12102} \end{APACrefDOI}
\PrintBackRefs{\CurrentBib}

\bibitem [\protect \citeauthoryear {%
Robbins%
\ \BBA {} Monro%
}{%
Robbins%
\ \BBA {} Monro%
}{%
{\protect \APACyear {1951}}%
}]{%
robbins1951stochastic}
\APACinsertmetastar {%
robbins1951stochastic}%
\begin{APACrefauthors}%
Robbins, H.%
\BCBT {}\ \BBA {} Monro, S.%
\end{APACrefauthors}%
\unskip\
\newblock
\APACrefYearMonthDay{1951}{}{}.
\newblock
{\BBOQ}\APACrefatitle {A stochastic approximation method} {A stochastic
  approximation method}.{\BBCQ}
\newblock
\APACjournalVolNumPages{The Annals of Mathematical
  Statistics}{22}{3}{400--407}.
\newblock
\begin{APACrefDOI} \doi{10.1214/aoms/1177729586} \end{APACrefDOI}
\PrintBackRefs{\CurrentBib}

\bibitem [\protect \citeauthoryear {%
Rupp%
, Templin%
\BCBL {}\ \BBA {} Henson%
}{%
Rupp%
\ \protect \BOthers {.}}{%
{\protect \APACyear {2010}}%
}]{%
rupp2010diagnostic}
\APACinsertmetastar {%
rupp2010diagnostic}%
\begin{APACrefauthors}%
Rupp, A\BPBI A.%
, Templin, J.%
\BCBL {}\ \BBA {} Henson, R\BPBI A.%
\end{APACrefauthors}%
\unskip\
\newblock
\APACrefYear{2010}.
\newblock
\APACrefbtitle {Diagnostic Measurement: Theory, Methods, and Applications}
  {Diagnostic measurement: Theory, methods, and applications}.
\newblock
\APACaddressPublisher{New York, NY}{Guilford Press}.
\PrintBackRefs{\CurrentBib}

\bibitem [\protect \citeauthoryear {%
Shoben%
}{%
Shoben%
}{%
{\protect \APACyear {1983}}%
}]{%
shoben1983applications}
\APACinsertmetastar {%
shoben1983applications}%
\begin{APACrefauthors}%
Shoben, E\BPBI J.%
\end{APACrefauthors}%
\unskip\
\newblock
\APACrefYearMonthDay{1983}{}{}.
\newblock
{\BBOQ}\APACrefatitle {Applications of multidimensional scaling in cognitive
  psychology} {Applications of multidimensional scaling in cognitive
  psychology}.{\BBCQ}
\newblock
\APACjournalVolNumPages{Applied Psychological Measurement}{7}{4}{473--490}.
\newblock
\begin{APACrefDOI} \doi{10.1177/014662168300700406} \end{APACrefDOI}
\PrintBackRefs{\CurrentBib}

\bibitem [\protect \citeauthoryear {%
Skager%
, Schultz%
\BCBL {}\ \BBA {} Klein%
}{%
Skager%
\ \protect \BOthers {.}}{%
{\protect \APACyear {1966}}%
}]{%
skager1966multidimensional}
\APACinsertmetastar {%
skager1966multidimensional}%
\begin{APACrefauthors}%
Skager, R\BPBI W.%
, Schultz, C\BPBI B.%
\BCBL {}\ \BBA {} Klein, S\BPBI P.%
\end{APACrefauthors}%
\unskip\
\newblock
\APACrefYearMonthDay{1966}{}{}.
\newblock
{\BBOQ}\APACrefatitle {The multidimensional scaling of a set of artistic
  drawings: Perceived structure and scale correlates} {The multidimensional
  scaling of a set of artistic drawings: Perceived structure and scale
  correlates}.{\BBCQ}
\newblock
\APACjournalVolNumPages{Multivariate Behavioral Research}{1}{4}{425--436}.
\newblock
\begin{APACrefDOI} \doi{10.1207/s15327906mbr0104_2} \end{APACrefDOI}
\PrintBackRefs{\CurrentBib}

\bibitem [\protect \citeauthoryear {%
Subkoviak%
}{%
Subkoviak%
}{%
{\protect \APACyear {1975}}%
}]{%
subkoviak1975use}
\APACinsertmetastar {%
subkoviak1975use}%
\begin{APACrefauthors}%
Subkoviak, M\BPBI J.%
\end{APACrefauthors}%
\unskip\
\newblock
\APACrefYearMonthDay{1975}{}{}.
\newblock
{\BBOQ}\APACrefatitle {The use of multidimensional scaling in educational
  research} {The use of multidimensional scaling in educational
  research}.{\BBCQ}
\newblock
\APACjournalVolNumPages{Review of Educational Research}{45}{3}{387--423}.
\newblock
\begin{APACrefDOI} \doi{10.3102/00346543045003387} \end{APACrefDOI}
\PrintBackRefs{\CurrentBib}

\bibitem [\protect \citeauthoryear {%
Takane%
}{%
Takane%
}{%
{\protect \APACyear {2006}}%
}]{%
takane200611}
\APACinsertmetastar {%
takane200611}%
\begin{APACrefauthors}%
Takane, Y.%
\end{APACrefauthors}%
\unskip\
\newblock
\APACrefYearMonthDay{2006}{}{}.
\newblock
{\BBOQ}\APACrefatitle {11 Applications of Multidimensional Scaling in
  Psychometrics} {11 applications of multidimensional scaling in
  psychometrics}.{\BBCQ}
\newblock
\APACjournalVolNumPages{Handbook of Statistics}{26}{}{359--400}.
\newblock
\begin{APACrefDOI} \doi{10.1016/S0169-7161(06)26011-5} \end{APACrefDOI}
\PrintBackRefs{\CurrentBib}

\bibitem [\protect \citeauthoryear {%
van~der Linden%
}{%
van~der Linden%
}{%
{\protect \APACyear {2008}}%
}]{%
van2008using}
\APACinsertmetastar {%
van2008using}%
\begin{APACrefauthors}%
van~der Linden, W\BPBI J.%
\end{APACrefauthors}%
\unskip\
\newblock
\APACrefYearMonthDay{2008}{}{}.
\newblock
{\BBOQ}\APACrefatitle {Using response times for item selection in adaptive
  testing} {Using response times for item selection in adaptive
  testing}.{\BBCQ}
\newblock
\APACjournalVolNumPages{Journal of Educational and Behavioral
  Statistics}{33}{1}{5--20}.
\newblock
\begin{APACrefDOI} \doi{10.3102/1076998607302626} \end{APACrefDOI}
\PrintBackRefs{\CurrentBib}

\bibitem [\protect \citeauthoryear {%
Wang%
, Zhang%
, Douglas%
\BCBL {}\ \BBA {} Culpepper%
}{%
Wang%
\ \protect \BOthers {.}}{%
{\protect \APACyear {2018}}%
}]{%
wang2018using}
\APACinsertmetastar {%
wang2018using}%
\begin{APACrefauthors}%
Wang, S.%
, Zhang, S.%
, Douglas, J.%
\BCBL {}\ \BBA {} Culpepper, S.%
\end{APACrefauthors}%
\unskip\
\newblock
\APACrefYearMonthDay{2018}{}{}.
\newblock
{\BBOQ}\APACrefatitle {Using Response Times to Assess Learning Progress: A
  Joint Model for Responses and Response Times} {Using response times to assess
  learning progress: A joint model for responses and response times}.{\BBCQ}
\newblock
\APACjournalVolNumPages{Measurement: Interdisciplinary Research and
  Perspectives}{16}{1}{45--58}.
\newblock
\begin{APACrefDOI} \doi{10.1080/15366367.2018.1435105} \end{APACrefDOI}
\PrintBackRefs{\CurrentBib}

\bibitem [\protect \citeauthoryear {%
Zhan%
, Jiao%
\BCBL {}\ \BBA {} Liao%
}{%
Zhan%
\ \protect \BOthers {.}}{%
{\protect \APACyear {2018}}%
}]{%
zhan2018cognitive}
\APACinsertmetastar {%
zhan2018cognitive}%
\begin{APACrefauthors}%
Zhan, P.%
, Jiao, H.%
\BCBL {}\ \BBA {} Liao, D.%
\end{APACrefauthors}%
\unskip\
\newblock
\APACrefYearMonthDay{2018}{}{}.
\newblock
{\BBOQ}\APACrefatitle {Cognitive diagnosis modelling incorporating item
  response times} {Cognitive diagnosis modelling incorporating item response
  times}.{\BBCQ}
\newblock
\APACjournalVolNumPages{British Journal of Mathematical and Statistical
  Psychology}{71}{2}{262--286}.
\newblock
\begin{APACrefDOI} \doi{10.1111/bmsp.12114} \end{APACrefDOI}
\PrintBackRefs{\CurrentBib}

\end{thebibliography}

\end{document}